# Anomalous Power Factor Enhancement and Local Structural Transition in Ni-Doped TiCoSb

Suman Mahakal[1], Pallabi Sardar[1, 2], Diptasikha Das[2], Subrata Jana[3], Swapnava Mukherjee[1, 2], Biplab Ghosh[4], Shamima Hussain[5], Santanu K. Maiti[6], and Kartick Malik*[1]

[1] *Department of Physics, Vidyasagar Metropolitan College, Kolkata-700006, India.*
[2]*Department of Physics, Adamas University, Kolkata-700126, India.*
[3]*Institute of Physics, Faculty of Physics, Astronomy and Informatics, Nicolaus Copernicus University, Torun, ul. Grudziadzka 5, 87-100 Torun, Poland.*
[4]*Beamline Development & Application Section, Bhabha Atomic Research Centre, Trombay, Mumbai-400 085, India*
[5] *UGC-DAE Consortium for Scientific Research, Kalpakkam Node, Kokilamedu 603 104, Tamil Nadu, India.*
[6] *Physics and Applied Mathematics Unit, Indian Statistical Institute, 203 Barrackpore Trunk Road, Kolkata-700 108, India.*

**ABSTRACT**

We report a significant enhancement (~269%) in the power factor (PF) and a local structural transition in Ni-doped TiCoSb samples ($TiCo_{1-x}Ni_xSb$, x= 0.0, 0.01, 0.02, 0.03, 0.04, and 0.06). First-principles calculations reveal that even minute Ni doping induces a substantial shift in the Fermi level ($E_F$) and alters the density of states (DOS). Structural analysis via Rietveld refinement of X-ray diffraction (XRD) data shows anomalous behavior at x = 0.02, supported by Williamson-Hall and modified methods. X-ray absorption spectroscopy (XAS) at the Ti and Co K-edges further confirms a pronounced local structural change at this composition. These structural transitions are consistent with temperature-dependent resistivity (ρ(T)) and thermopower (S(T)) data, which reflect changes in $E_F$ and disorder. Analysis of Lorentz number and scattering parameters reinforces the observed modifications in the electronic structure. The simultaneous enhancement of S and electrical conductivity at x = 0.02 is attributed to the disorder-to-order transition, leading to the marked rise in PF.

## 1. Introduction

Conversion of abundant waste heat into electricity using thermoelectric (TE) material has the potential to address the global energy demand in the future. The dimensionless parameter ZT [$(=S^2\sigma/\kappa)T$], defined by Seebeck coefficient (S), electrical conductivity (σ), and thermal conductivity (κ) at temperature T, is a key factor that determines the efficiency of a TE device module [1, 2]. $S^2\sigma$ is known as the TE Power Factor (PF). The performance of a device module is directly correlated with output power density (ω) or power conversion efficiency in certain ambient environments. The relation between average power factor ($PF_{ave}$) and ω is given by [3, 4]

$$\omega = \frac{(T_h - T_c)^2}{4l} \mathrm{PF}_{\mathrm{ave}} \quad (1)$$

when a TE material of length $l$ is kept at a temperature difference $T_h$-$T_c$. However, a TE module in the presence of an abundant supply of waste heat may be more economically viable due to the high $PF_{ave}$ of the TE element [4].

TiCoSb is a typical p-type semiconducting material, drawing attention as a potential mid-temperature TE material [5] owing to non-toxicity, good electrical properties, high thermal stability, and robust mechanical properties. TiCoSb crystallizes with an MgAgAs-type structure. TiCoSb, a half-Heusler (HH) compound with an 18-valence-electron count (VEC), has potential for TE applications owing to a narrow band gap and sharp slope in density of states (DOS) near the Fermi level [5]. However, efficiency is limited due to a high κ of the material [6]. Efforts are employed to enhance the TE properties of TiCoSb-based semiconductors through simultaneous doping and alloying at various sites [7, 8]. Alloying reduces κ, whereas electrical properties are optimized by doping [9-12]. Theoretical calculations show that conduction band maxima and valence band minima are constituted of the contribution from Ti and Co orbitals [13, 14]. Romaka et al. observed that Ni doping in TiCoSb alloy causes a shift in the position of the Fermi energy ($E_F$) in the theoretically estimated DOS [15]. Yan et al. reported that Ti substitution in $Hf_{1-x}Ti_xCoSb_{0.8}Sn_{0.2}$ causes enhancement in PF due to manipulation of carrier concentration [16]. However, PF values may be optimized up to a certain extent due to the inverse relation between S and σ [17, 18]. Wang et al. reported that Co doping in Ti(Fe, Co, Ni)Sb causes enhancement in PF due to simultaneous increase in carrier concentration and density of states effective mass [19]. However, decoupling of S and σ is another route to enhance PF [7]. Van Du et al. mentioned that Sb doping at the Sn site of $Ti_{0.5}Zr_{0.5}NiSn$ alloy enhances σ without significant degradation of S, resulting in a high PF [20]. Further, Ta doping at the Ti site in TiCoSb also increases PF by decoupling the σ and S [7]. Yang et al. extensively studied the effect of Bi doping on the structure and TE properties of $NaCoO_2$ and reported an enhancement in PF [21]. Further, doping has a significant influence on crystal structure and electronic properties, which may be employed to enhance PF. There are some reports on the effect of Ni doping in TE, and the electronic properties of TiCoSb, and a peculiar behavior is observed for high Ni doping [22-24]. However, to the best of our knowledge, there are very limited reports on the simultaneous enhancement of σ and S of TiCoSb-based TE material due to small Ni doping.

The purpose of the study is to enhance the PF of TiCoSb by simultaneously increasing the σ and S. Further, efforts have been employed to reveal the correlation between structural change and transport properties due to a minute amount of Ni doping in TiCoSb TE material. Very high enhancement ~269% in PF is observed in $TiCo_{1-x}Ni_xSb$ (x=0.0, 0.01, 0.02, 0.03, 0.04, and 0.06) synthesized samples. The effect of small Ni doping on the electronic structure is studied by first principle density functional theory (DFT). Long-range structural information of synthesized samples is revealed by X-ray diffraction (XRD). X-ray absorption spectroscopy (XAS) is employed to explore the change in local structural arrangements. PFs are estimated from the experimental data of S(T) and temperature-dependent ρ(T). The key findings of the present study are: (i) a minute amount of Ni doping in TiCoSb induces a shift in the Fermi level $E_F$, leading to a transition from semiconducting to metallic behavior, as evidenced by both theoretical and experimental studies. (ii) Local structural rearrangement in $TiCo_{1-x}Ni_xSb$ ($0 \leq x \leq 0.6$) samples with an anomaly at x=0.02 is observed. (iii) Experimental S(T) and ρ(T) data are corroborated with long-range structural data, and a shift in $E_F$ due to the replacement of Co by Ni in the TiCoSb matrix is revealed. (iv) Simultaneous increase in ρ and S results enhancement in PF ~269%, accompanied by changes in estimated local and long-range structural parameters.

The article is organised as follows. Section II provides the detailed procedure of sample synthesis and characterization methods employed to analyze the experimental data. Computational details are presented in Section III. Theoretical analysis and experimental results are critically discussed in Section IV. The essential results are summarised in Section V.

## 2. Experimental Details

High-purity Ti, Co, Ni, and Sb (99.999% purity, Alfa Aesar, UK) with weighted stoichiometry were arc melted to synthesize $TiCo_{1-x}Ni_xSb$ (x=0.0, 0.01, 0.02, 0.03, 0.04, and 0.06) HH polycrystalline samples. The melting process was carried out using a vacuum arc furnace and repeated three times in the presence of inert gas flow (Ar) to ensure homogeneity and avoid oxidation during the melting of metals. Arc-melted ingots were vacuum sealed in quartz tubes under $10^{-3}$ Pa pressure to prevent oxidation during soaking at high temperature. The sealed quartz ampules underwent annealing at 1173 K for 80 hours and were subsequently cooled to room temperature at a rate of 10 K/h.

Room-temperature powder X-ray diffraction (XRD) (Model: X'Pert Powder, PANalytical) measurement was carried out using Cu-Kα radiation (λ~ 0.15418 nm) for the long-range structural characterization of the synthesized samples. XRD data were taken in the range $20° < 2\theta < 80°$ in θ-2θ geometry. Rietveld refinement of XRD data using FullProf software was employed to carry out in depth structural characterization [25, 26]. The Williamson-Hall and its modified methods [27-29] were utilized to evaluate lattice strain (ε), crystalline size (d) and relative dislocation densities ($N_D/ N_{D|x=0.0}$) of the synthesized $TiCo_{1-x}Ni_xSb$ samples using the XRD data.

XAS measurements were performed at Co and Ti K-edge in transmission and fluorescence mode, respectively. However, due to the limitation in operational energy range (4 keV to 25 keV), data collection at the Sb K-edge was not feasible. XAS spectra of standard Co and Ti metal foils were utilized to calibrate the spectra of the synthesized samples at the Co and Ti K-edge.

Temperature-dependent resistivity ρ(T) measurements of the synthesized polycrystalline alloys were carried out using the conventional four probe method down to 10 K. Temperature-dependent S(T) measurements were performed down to 20K, employing the standard differential technique [30].

## 3. Computational details

A theoretical study of $TiCo_{1-x}Ni_xSb$ HH alloys was conducted using first principles density functional theory (DFT) as deployed in the Vienna Ab initio Simulation Package (VASP) [31]. It computed the pseudopotential of elements using a projected augmented wave technique. The Perdew−Burke−Ernzerhof (PBE) functional, within the framework of the generalized gradient approximation (GGA), was used as the electron exchange-correlation function [32]. In order to optimise the kinetic energy cut-off, several calculations were performed, varying the values from 100 eV to 600 eV. Finally, the kinetic energy cut-off of 520 eV and 10 x 10 x 10 k-mess sampling of the first Brillouin zone (BZ) were used for the expansion of plane-wave basis sets in the Monkhorst-Pack (MP) mesh for the geometry optimization. The Gaussian smearing approach, with a smearing width of 0.01 eV, was employed to integrate the BZ. The total energy and force convergence criteria of the calculation were set to $10^{-6}$ eV and $10^{-3}$ eVÅ$^{-1}$, respectively. A primitive unit cell with one Ti, Co, and Sb atom was used. Both the ionic sites and cell parameters were relaxed during the optimization process. Throughout the present study, all calculations were performed using this optimised crystal structure. To create the Ni doped system, a conventional unit cell, consisting of twelve atoms (four Ti, four Co, and four Sb), was multiplied by 2 × 2 × 2. Thereafter, one and two Ni atoms were substituted at the Co site, leading to Ni concentration of ~3% and ~6%, respectively.

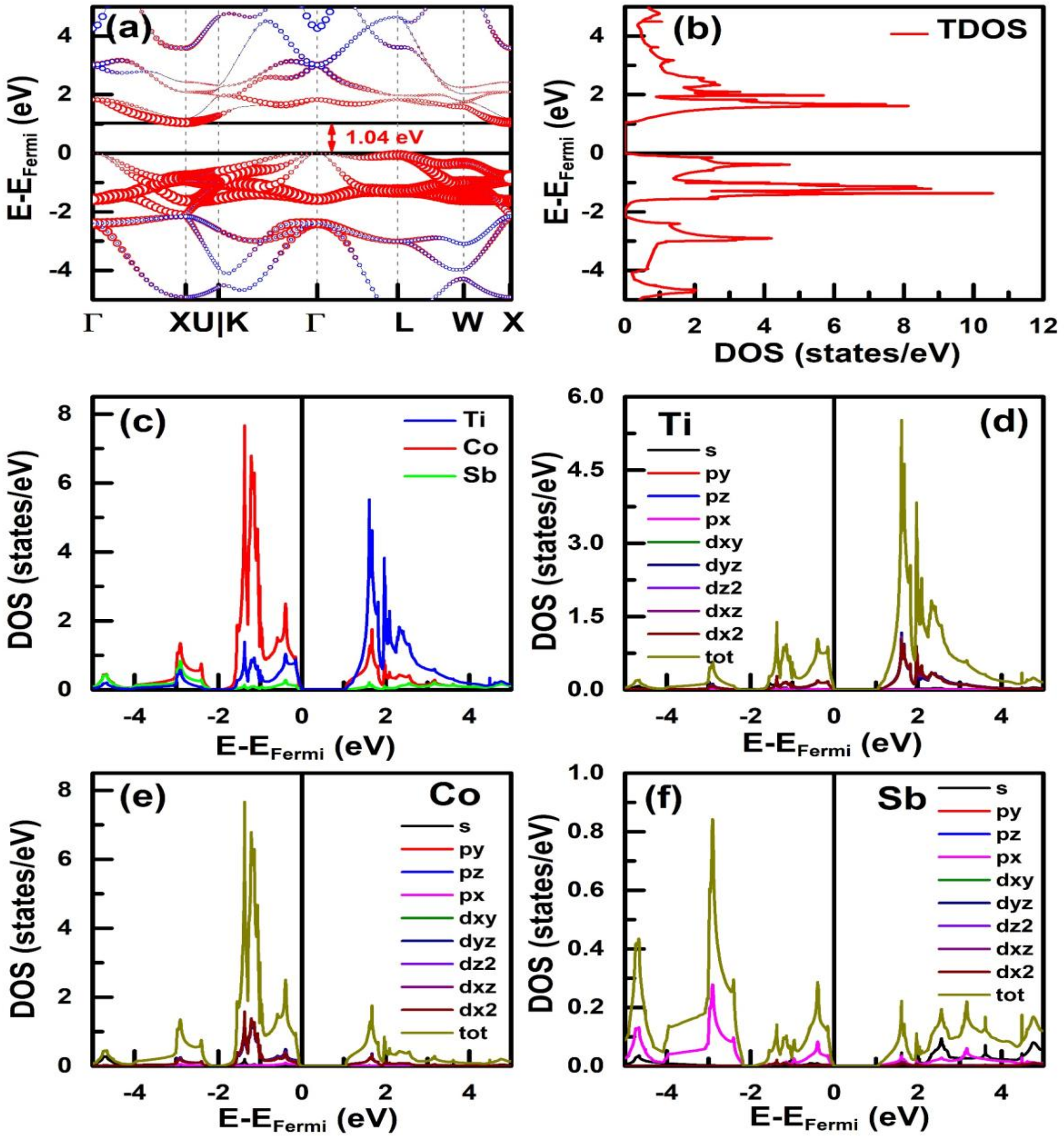

**Fig. 1**. **(Color online).** Theoretical illustration of TiCoSb compound, includes: (a) projected band structure (contribution from Ti, Co, and Sb atoms is represented by blue, red, and green color of bubbles, respectively), (b) total density of states (TDOS), (c) combined partial density of states (PDOS); and PDOS for (d) Ti, (e) Co and (f) Sb.

## 4. Results and Discussion

### *4.1. Theoretical analysis*

TiCoSb is classified as a non-magnetic semiconducting material amid the HH compounds [1, 33]. The compound, TiCoSb, is in the category of XYZ structure with 1:1:1 stoichiometry and crystallized as MgAgAs type structure with space group $\mathbf{F\bar{4}3m}$ [1, 34]. It is crucial to mention that three different atomic orderings exist for the MgAgAs type HH structure [1]. The crystal structural optimisation is performed for all three possible orderings with different Wyckoff positions of Ti, Co and Sb atoms. The TiCoSb HH compound is found to be energetically more favourable for the atomic configuration where the Ti, Co, and Sb atoms occupy the 4b (0.5, 0.5, 0.5), 4c (0.25, 0.25, 0.25) and 4a (0, 0, 0) Wyckoff positions, respectively. The estimated value of the lattice parameter ($a$) of the primitive unit cell is 4.16 Å, i.e., for the conventional cell, $a$ is 5.89 Å. The $a$ value, estimated for the most stable structure of TiCoSb, exhibits good agreement with existing literature, [15] confirms the reliability of our calculations. The most stable configuration of the TiCoSb HH compound is considered for further calculations of electronic properties.

Figure 1(a) depicts the band structure of TiCoSb at various high symmetry points within the first Brillouin zone, calculated by employing ab initio density functional theory (DFT). The total and angular momentum projected partial density of states of the TiCoSb compound are also investigated. In order to establish a reliable framework of the analysis, the position of the Fermi energy is shifted to the zero reference point. The contributions of different atoms to the band states in Fig. 1(a) are represented using the bubbles of various colours. The size of each bubble signifies its weighted contribution. Figure 1(a) indicates that the TiCoSb compound shows p-type semiconducting behaviour with an indirect bandgap of approximately 1.04 eV. The valence band maximum (VBM) and the conduction band minimum (CBM) are at the Γ and X-point in the Brillouin zone, respectively. It is crucial to note that the valence band is primarily constituted from the contribution of Co orbitals (blue bubble), whereas Ti orbitals (red bubble) predominate in the conduction band. Careful investigatiion reveals that the bottom of the conduction band at X-point has a contribution of the Ti atoms, while the top of the valence band at Γ-point has a complex contribution of the Ti and Co orbitals. This orbital interplay at the Γ-point indicates strong hybridization between Ti and Co orbitals, which can influence the carrier dynamics.

Further scrutiny and in-depth study of the electronic structure are carried out employing the total density of states (TDOS) and partial density of states (PDOS), as illustrated in Fig. 1(b-f). The modified tetrahedron approach is used to estimate the DOS from Kohn-Sham eigenvalues on a highly dense k-grid. Figure 1(c) depicts the different atomic contributions to the total density of states, indicating that the electronic states primarily originate from the Co, Ti, and a minor contribution from Sb atoms. Figure 1(d-f) demonstrates the contributions to the TDOS from the different orbitals of Ti, Co, and Sb, respectively. It is notable that the dominant contributions in the DOS primarily come from Ti-3d and Co-3d orbitals, as shown in Fig. 1(d) and Fig. 1 (e), while Fig. 1(f) highlights that the Sb-5s and Sb-5p states mostly contribute to the electronic state of Sb.

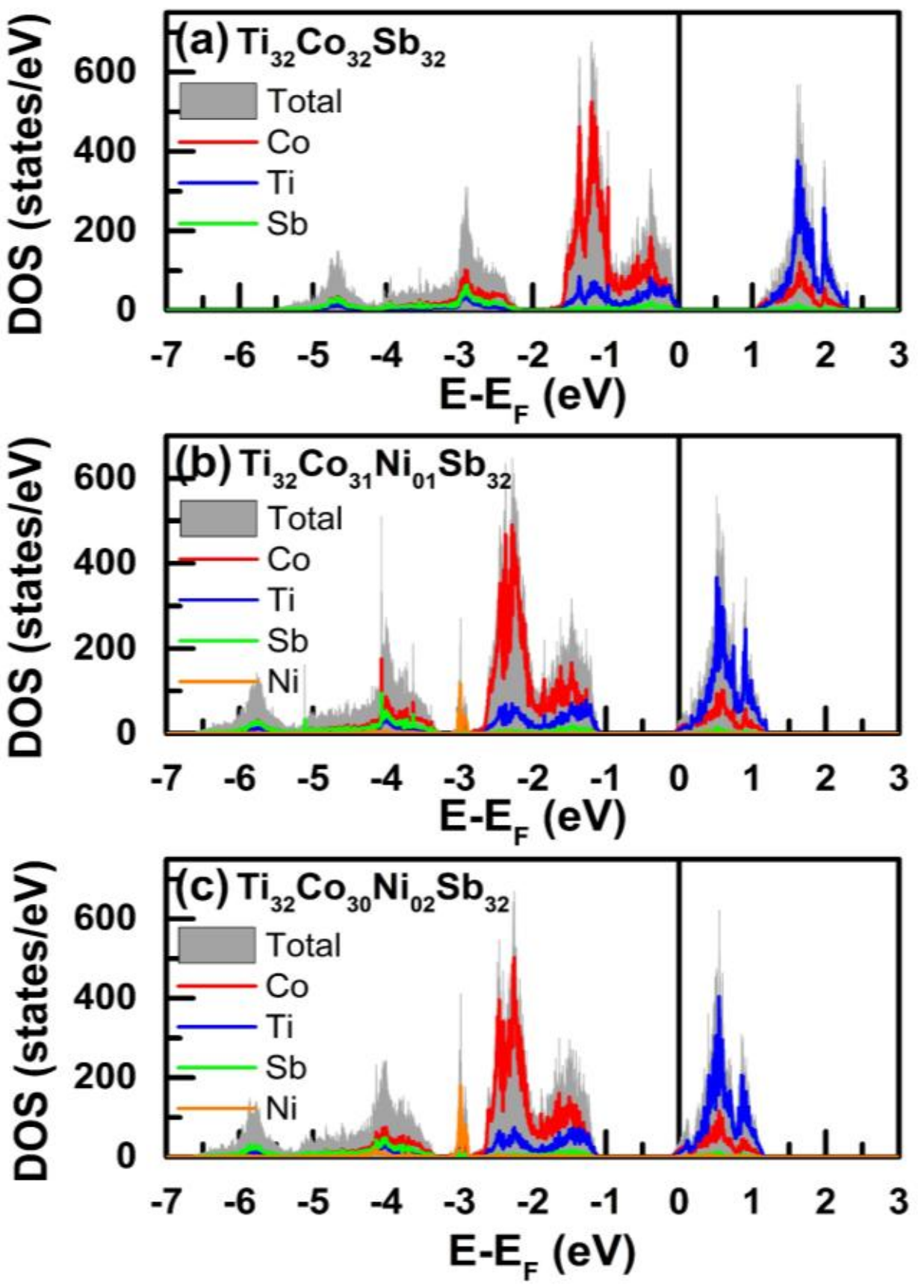

**Fig. 2. (Color online).** Density of states of (a) $Ti_{32}Co_{32}Sb_{32}$, (b) $Ti_{32}Co_{31}Ni_{1}Sb_{32}$, and (c) $Ti_{32}Co_{30}Ni_{2}Sb_{32}$ supercell, corresponding to the 0%, 3% and 6% Ni doping concentration, respectively.

To delve into the effect of Ni doping on the electronic structure of the $TiCo_{1-x}Ni_xSb$ compounds, a 2×2×2 supercell of the conventional unit cell of TiCoSb is utilized. We calculate the DOS of $Ti_{32}Co_{32}Sb_{32}$, $Ti_{32}Co_{31}Ni_{01}Sb_{32}$ and $Ti_{32}Co_{30}Ni_{02}Sb_{32}$ systems to achieve 0%, 3%, and 6% Ni doping levels, respectively. Other doping levels are avoided due to the high computational cost. Figure 2 demonstrates the density of states of the $Ti_{32}Co_{32}Sb_{32}$, $Ti_{32}Co_{31}Ni_{01}Sb_{32}$ and $Ti_{32}Co_{30}Ni_{02}Sb_{32}$ systems. It is worth noting that the Ni doping in the $TiCo_{1-x}Ni_xSb$ compound pushes the Fermi level into the conduction band without affecting the band gap, resulting in metallic behaviour for 3% and 6% Ni doping.

*4.2. Structural Characterization*

Figure 3 represents the XRD pattern of the synthesized $TiCo_{1-x}Ni_xSb$ ($0.00 \leq x \leq 0.06$) HH samples. All the synthesized samples are crystallized as MgAgAs type structure with space group $F\bar{4}3m$ [1]. Indistinguishable shift in position of the highest intense peak (220) with increasing

Ni concentration is observed due to the comparable atomic radius of Co and Ni [22]. Crystalline strain (ε) and average crystalline size (D) are estimated from XRD data, employing the Williamson-Hall equation [27]

$$\beta\cos\theta = \frac{k_B\lambda}{D} + 4\varepsilon\sin\theta. \quad (2)$$

β, $k_B$ and λ are the broadening in diffraction peak of the synthesized sample, the Boltzmann constant, and the wavelength of X-ray, respectively. Figure 4(a) illustrates the variation in ϵ and D as a function of Ni concentration. Negative ε of the synthesized samples indicates a compressive nature. ε of the synthesized $TiCo_{1-x}Ni_xSb$ samples decreases for $x \leq 0.02$ and enhances in the range of $0.02 < x \leq 0.06$, whereas D shows the opposite nature with increasing Ni concentration. Further, XRD data of the synthesized smples are employed to estimate dislocation densities ($N_D$) in $TiCo_{1-x}Ni_xSb$ matrix, using modified Williamson-Hall method [28, 29]. The detailed theory, fittings (Fig. S1), and calculations are presented in the supporting information (SI). Figure 4(b) represents the relative dislocation density ($N_D/N_D|_{x=0}$) i.e., change in dislocation density with respect to the dislocation density in synthesized TiCoSb (x=0) sample. It is crucial to note that the minimum $N_D$ is obtained for $TiCo_{0.98}Ni_{0.02}Sb$,

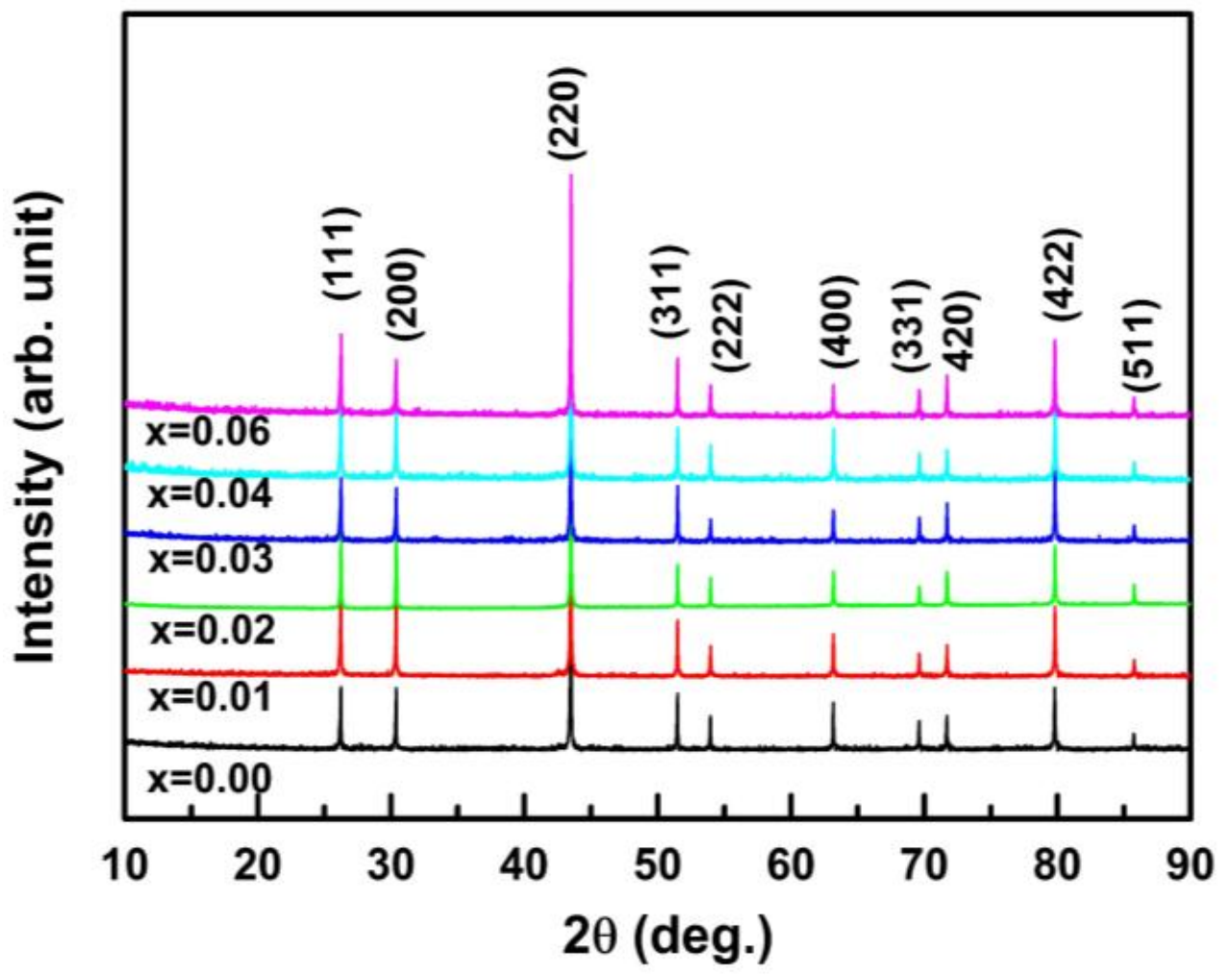

**Fig. 3. (Color online).** Room Temperature X-ray diffraction (XRD) of synthesized $TiCo_{1-x}Ni_xSb$ (x=0.0, 0.01, 0.02, 0.03, 0.04, 0.06) HH alloy using Cu-$K_\alpha$ source.

amid the synthesized samples.

In order to estimate structural parameters, experimental XRD data are fitted employing the Rietveld refinement technique, and the estimated parameters along with fitted graphs are shown in Fig. S2 of SI. A lower value of R-factor and Goodness of Fit ($\chi^2$) signifies a promising fit of the XRD data during Rietveld refinement. A trivial variation in $a$ for synthesized $TiCo_{1-x}Ni_xSb$ samples is observed, and $a$~0.5883 nm is estimated for TiCoSb sample from refinement, which is comparable with the optimised lattice parameter obtained from first principle ($a$=0.59 nm). A similar result for TiCoSb ($a$=0.5884 nm) is also reported by Webster and Ziebeck [35]. Figure 5(a) represents the non-monotonic variation of unit cell volume ($V$) of synthesized $TiCo_{1-x}Ni_xSb$ alloys as a function of Ni concentration (x). A decrease in $V$ with Ni doping is observed owing to slightly lower atomic radii of Ni (0.1620 nm) compared to Co (0.1670 nm), indicating successful replacement of Co by Ni atoms [22]. Mix-phase Rietveld refinement reveals the presence of CoTi embedded phases in the $TiCo_{1-x}Ni_xSb$ matrix, and the variation of weight percentage (wt%) of the host phase along with CoTi phases as a function of Ni concentration (x) are depicted in Fig. 5(b). Embedded phases, especially CoTi phases in TiCoSb HH matrix, is frequently observed due to the large difference in the melting point of the constituent elements [14, 36]. Incorporation of CoTi phase during mix-phase Rietveld refinement analysis provides an excellent fit, and optimum $\chi^2$ is achieved. It is important to note that maximum host phase is obtained for x=0.02. Further, the Debye-Waller factor (atomic displacement parameter, $B_{iso}$) of the synthesized samples, obtained after refinement, is presented in Fig. 5(c). Minimum $B_{iso}$ is observed for x=0.02 and increases on either side in Fig. 5(c), i.e., for $0.00 \leq x \leq 0.02$ and $0.02 \leq x \leq 0.06$. The difference in atomic size of Co and Ni may be a reason for the increase in positional disorder or atomic dislocation [36]. Further, the presence of embedded phases is also responsible for lattice dislocation [14, 36]. Increase in doping concentration and embedded phases for $x \geq 0.02$, results abrupt increase in $B_{iso}$.

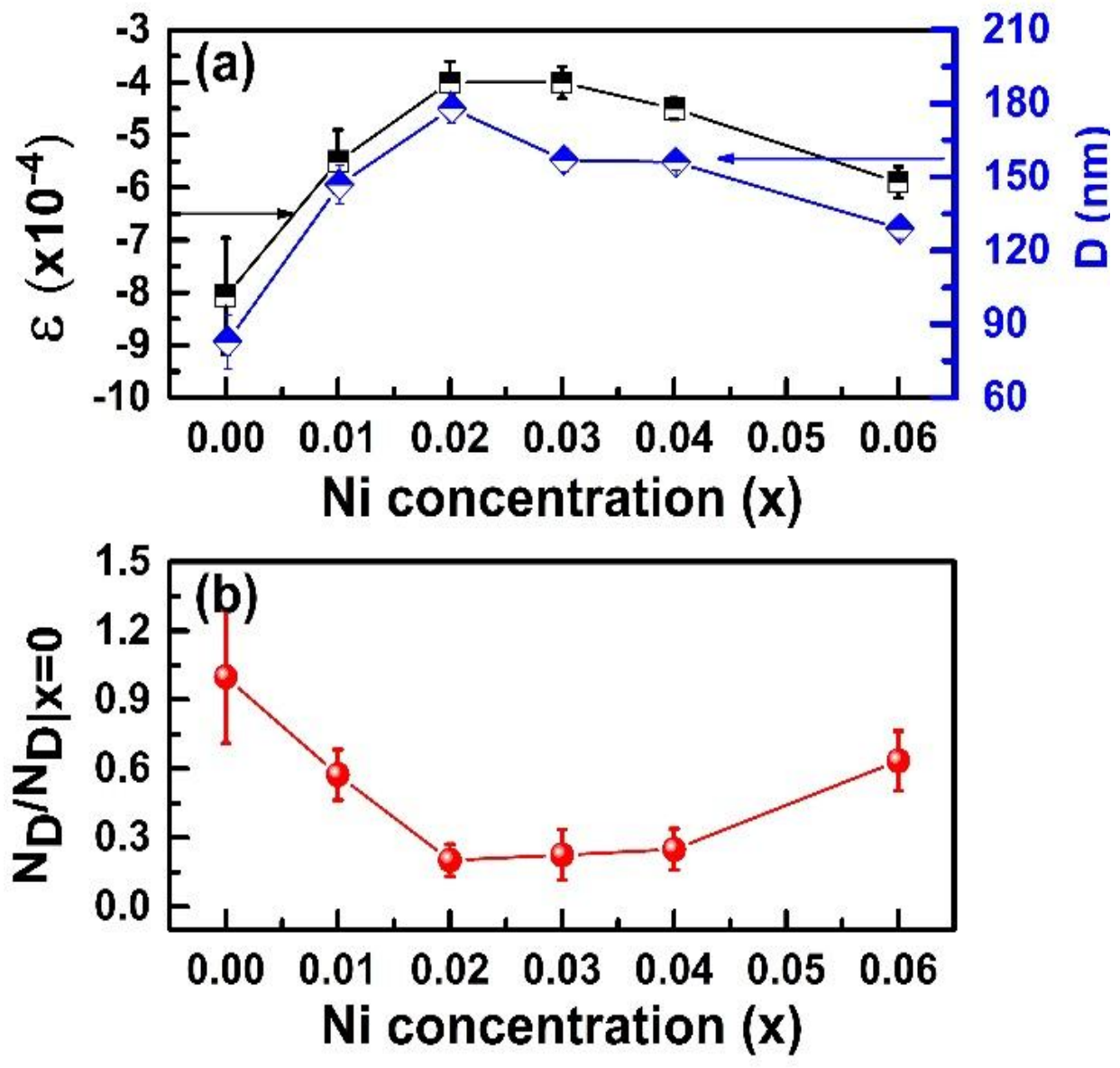

**Fig. 4. (Color online).** (a) Variation of lattice strain (ε) and crystalline size (D) with Ni concentration obtained from X-ray diffraction data using Williamson-Hall equation, (b) Ni concentration dependent relative dislocation density ($N_D/N_{D|x=0}$) estimated from X-ray diffraction data using modified Williamson-Hall equation of synthesized $TiCo_{1-x}Ni_xSb$ (x=0.0, 0.01, 0.02, 0.03, 0.04, 0.06) HH polycrystalline materials.

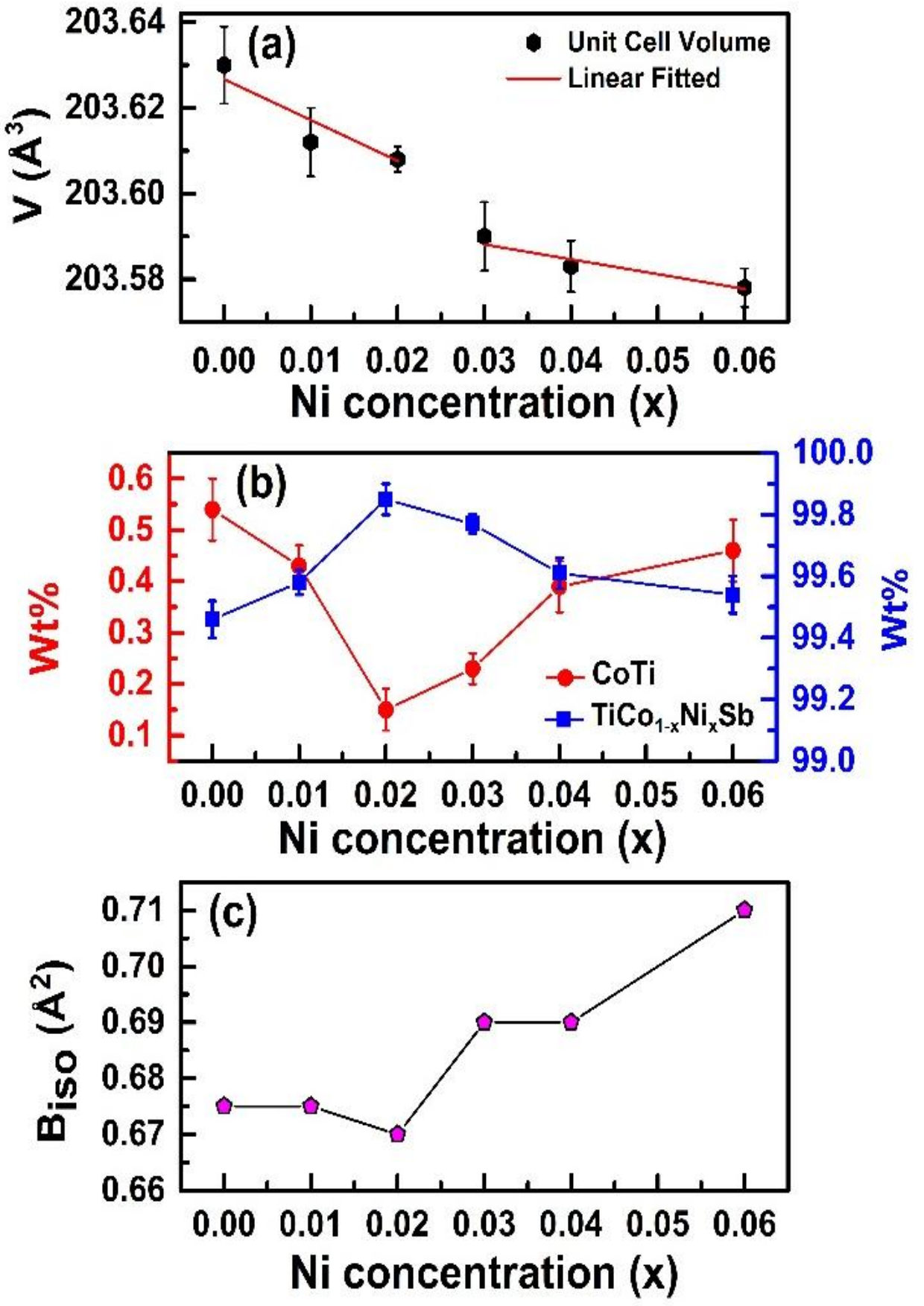

**Fig. 5. (Color online).** Variation of (a) unit cell volume, (b) wt% of host phase and embedded phase, and (c) Debye-Waller factor ($B_{iso}$) with Ni concentration of synthesized $TiCo_{1-x}Ni_xSb$ HH alloy. The red line in (a) indicates tead line.

It is crucial to note that atypical behaviour is observed in structural parameters, obtained by Rietveld refinement and analysis of XRD data for the $TiCo_{0.98}Ni_{0.02}Sb$ sample. Maximum host phase and minimum embedded phase, i.e., CoTi, is observed for x=0.02 (Fig. 5(b)), correlated with the variation in ε and D (Fig. 4(a)). Increase in ε of the synthesized samples may be related to lattice mismatch owing to the embedded phases [14]. Distortion at the grain boundary also has an impact on D. Yield strength ($\sigma_{GB}$), i.e., maximum stress required for plastic deformation, and average grain size (D), are related by the Hall-Petch formula [37, 38],

$$\sigma_{GB} = k_y D^{-1/2} \tag{3}$$

where $k_y$ is a material-dependent constant. The low strength ($\sigma_{GB}$) due to lattice mismatch or dislocation at the grain boundary causes a decrease in D [14]. It is important to mention that the highest D value for x=0.02 (Fig. 4(a)) attributes to the minimum dislocation due to embedded phases amid the synthesized $TiCo_{1-x}Ni_xSb$ (x=0, 0.01, 0.02, 0.03, 0.04, and 0.06) samples. The least value of $N_D$ also corroborates to the result. However, a comparable, better fitting for the $TiCo_{0.98}Ni_{0.02}Sb$ sample (Fig. S3 of SI) to estimate D and ε, employing the Williamson-Hall method indicates improvement in long-range structural order. Further, careful investigation of Fig. 4 reveals a sudden drop associated with slope change at x=0.02 for the synthesized $TiCo_{1-x}Ni_xSb$ alloys. Rietveld refinement results indicate an anomaly for x=0.02, and long-range structural disorder is correlated with a change in *V* and wt% of embedded phases.

### *4.3. Local structure characterization using XAS*

The effect of Ni doping on the local structure of the synthesized $TiCo_{1-x}Ni_xSb$ (x=0, 0.01, 0.02, 0.03, 0.04, and 0.06) samples is explored by XAS, using synchrotron radiation [39]. XAS data are taken at the Co and Ti K-edge.

**Table 1.** Degeneracy (N), scattering length ($R_{eff}$), rank, and types of scattering are presented for all probable scattering paths at Co K-edge and Ti K-edge of the synthesized $TiCo_{1-x}Ni_xSb$ (x=0.0, 0.01, 0.02, 0.03, 0.04, and 0.06) HH samples, used during FEFF modelling. Highlighted paths are included in the analysis during fitting.

| Co K-edge | | | | | Ti K-edge | | | | |
|---|---|---|---|---|---|---|---|---|---|
| Scattering Path | N | $R_{eff}$ (Å) | Rank | Type of Scattering | Scattering Path | N | $R_{eff}$ (Å) | Rank | Type of Scattering |
| **@ Sb8.1 @** | **4.00** | **2.552** | **100** | **Single** | **@ Co7.1 @** | **4.00** | **2.536** | **100** | **Single** |
| **@ Ti3.1 @** | **4.00** | **2.552** | **80.09** | **Single** | **@ Sb9.1 @** | **6.00** | **2.928** | **100** | **Single** |
| @ Sb8.1 Ti3.1 @ | 24.00 | 4.025 | 10.36 | Other double | **@ Co7.1 Sb9.1 @** | **24.00** | **4.000** | **10.86** | **Other double** |
| **@ Co5.1 @** | **12.00** | **4.167** | **64.65** | **Single** | **@ Ti3.1 @** | **12.00** | **4.141** | **63.88** | **Single** |
| @ Sb8.1 Sb10.1 @ | 12.00 | 4.636 | 6.63 | Other double | @ Co7.1 Co6.1 @ | 12.00 | 4.607 | 4.07 | Other double |
| @ Sb8.1 Co5.1 @ | 24.00 | 4.636 | 20.31 | Other double | @ Co7.1 Ti3.1 @ | 24.00 | 4.607 | 15.16 | Other double |
| @ Ti3.1 Ti0.1 @ | 12.00 | 4.636 | 7.55 | Other double | **@ Co7.2 @** | **12.00** | **4.856** | **40.75** | **Single** |
| **@ Ti3.1 Co5.1 @** | **24.00** | **4.636** | **20.69** | **Other double** | @ Sb9.1 Sb11.1 @ | 24.00 | 4.999 | 5.83 | Other double |
| **@ Sb8.2 @** | **12.00** | **4.887** | **63.27** | **Single** | @ Sb9.2 Ti2.1 @ | 48.00 | 4.887 | 19.77 | Other double |
| **@ Ti3.2 @** | **12.00** | **4.887** | **51.26** | **Single** | | | | | |

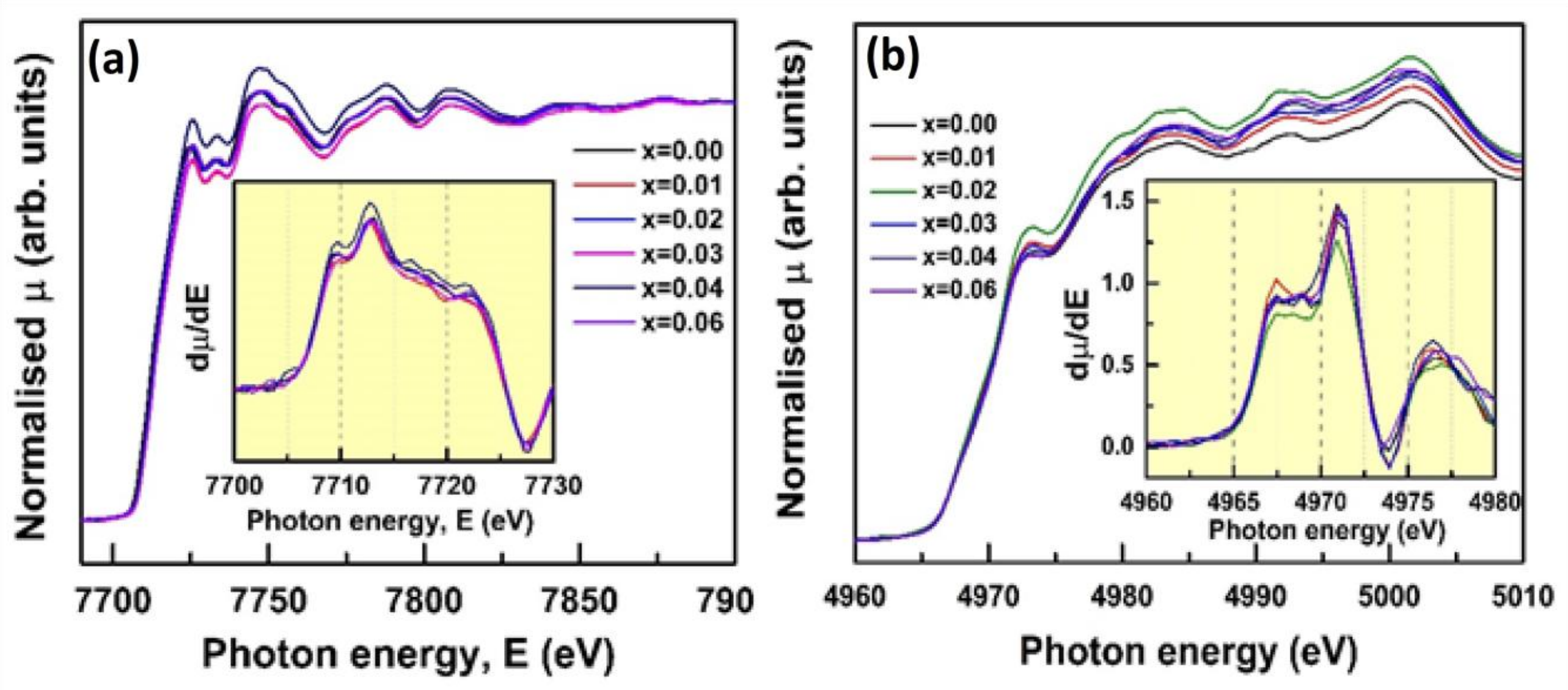

**Fig. 6. (Color online).** Normalised XENES spectra of synthesized $TiCo_{1-x}Ni_xSb$ at (a) Co K-edge and (b) Ti K-edge. Insets in (a) and (b) show the variation of 1$^{st}$ derivative of the edge jump.

XAS data reveal electronic features as well as local structural periodicity of the synthesized HH alloys [40]. Two regions of XAS data, (a) X-ray absorption near edge structure (XANES) and (b) extended X-ray absorption fine structure (EXAFS) for Co K-edge and Ti K-edge are extensively analyzed. Co and Ti K-edge spectra are obtained due to the ejection of an electron from the core level of the atom using X-ray energy, obtained from synchrotron source. Schematic diagrams for the transition of electrons in Co and Ti atoms are presented in Fig. S4 of SI. According to the electronic configuration of Co and Ti atoms in TiCoSb HH alloy (Co: [Ar] $3d^94s^0$ and Ti: [Ar] $3d^24s^2$), two types of transitions are allowed. The XANES spectrum of Co and Ti K-edge is obtained due to the excitation of 1s electrons to the unoccupied 4p level as the final state. The details of possible and forbidden transitions are discussed in the SI. Transitions from an occupied to a continuum state result in the EXAFS spectrum. The XAS data of the synthesized samples are given in Figs. S5 and S6 of SI. Energy dependent absorption coefficient, $\mu(E) \propto \ln(I_0/I_{t/f})$ ($I_0$ is the intensity of incident X-ray and $I_{t/f}$ is the intensity of the transmitted (or fluorescence-detected) signal) for the Co and Ti K-edge spectra is transformed into the fine structure function $\chi(k)$ [41]. The $k^2\chi(k)$ functions are Fourier transformed to real space (R-space), employing the ARTEMIS software package [42], and $k^2\chi(R)$ versus R data are fitted to reveal the interaction with the nearest neighbor. The quality of fittings is determined by $R_f$ [41, 43]. The details of XAS data and intermediate steps of data processing for synthesized $TiCo_{1-x}Ni_xSb$ alloys are presented in the SI. However, the scattering paths, included during fittings of R-space data for both Co and Ti K-edge, are given in Table 1, and reasons for excluding other paths are explained in SI.

The XANES spectra of synthesized $TiCo_{1-x}Ni_xSb$ HH alloys at Co and Ti K-edge are given in Fig. 6(a) and Fig. 6(b), respectively. The first derivative of the edge jump of XANES spectra is depicted in the inset of Fig. 6(a) and Fig. 6(b). Ti K-edge data (Fig. 6(b)) divulge no shift, indicating a small amount of Ni doping does not affect the environment and oxidation states of the synthesized samples. Figure 6(a) reveals negligible shift with increasing Ni concentration (x) in synthesized $TiCo_{1-x}Ni_xSb$ samples. The result has two aspects, Ni atoms replace Co atoms with similar oxidation states, and the average change in oxidation state is negligible due to minute doping of Ni atoms in TiCoSb HH alloy [44]. However, substitution of Co by Ni atoms donates one excess valence electron to the TiCoSb system and concomitantly changes oxidation states of Co and Ti in TiCoSb HH alloy [15]. In the present study, a minute amount of Ni doping ($0 \leq x \leq 0.06$) at the Co site causes an insignificant change in oxidation states of the synthesized $TiCo_{1-x}Ni_xSb$ alloys.

**Table 2.** R-factor and Happiness of fit values obtained after the fitting of EXAFS data using ARTEMIS software at the Co and Ti K-edge for the synthesized $TiCo_{1-x}Ni_xSb$ alloys, varying Ni concentrations.

| Ni Concentration | Co K-edge | | Ti K-edge | |
|---|---|---|---|---|
| (x) | R-factor | Happiness of fit | R-factor | Happiness of fit |
| 0.00 | 0.01576 | 97.00 | 0.01926 | 91.67 |
| 0.01 | 0.01384 | 96.70 | 0.01429 | 93.70 |
| 0.02 | 0.00498 | 94.51 | 0.03933 | 84.25 |
| 0.03 | 0.02111 | 96.10 | 0.01668 | 85.40 |
| 0.04 | 0.02287 | 93.98 | 0.02075 | 85.47 |
| 0.06 | 0.01826 | 98.34 | 0.02029 | 79.43 |

The $k^2$ weighted $\chi(k)$, $k^2\chi(k)$ and corresponding fittings, using theoretical modelling for Co K-edge EXAFS spectrum of the synthesized $TiCo_{1-x}Ni_xSb$ (x=0, 0.01, 0.02, 0.03, 0.04, and 0.06) samples are given in SI (Fig. S7). Figure 7(a) represents $k^2\chi(R)$ vs. R plot and theoretical fittings employing ARTEMIS software [42]. The Fourier transformed data show well-defined peaks at around 2.5 Å and 3.75 Å. Theoretical modelling of the data suggests that the peak at ~2.5 Å is due to the single scattering in the coordination shell, and the peak at ~3.75 Å is related to scattering owing to the second nearest neighbour shell. However, careful observation reveals a splitting in the first prominent peak (~2.5 Å). The energy of the illuminated X-ray is in the Co K-edge, and Co is tetrahedrally coordinated with four Sb and four Ti atoms in TiCoSb HH alloy [1, 45]. Peaks due to scattering from tetrahedrally coordinated Ti and Sb atoms may arise in the same position, but a difference in scattering phases of Ti and Sb atoms causes splitting. Theoretical modelling reveals that peaks at ~2.5 Å are related to Co-Ti, Co-Sb bonds, and the peak at ~3.75 Å arises due to dodecahedral bonds by 12 Co atoms in the TiCoSb system [45]. The contributions owing to other multiple scatterings are reflected as small peaks in the regions between the three main peaks related to single scattering. The negligible difference between theoretical and experimental data of EXAFS for the synthesized $TiCo_{1-x}Ni_xSb$ HH alloys, both in k-space and R-space, signifies the reliability of the fittings. However, small values of R-factors, presented in Table 2, quantify the reliability of good fits [41, 43]. The estimated bond lengths (R) for Co-Sb, Co-Ti, and Co-Co bonds of the synthesized samples are depicted in Fig. 7(b). The disorder parameter ($\sigma^2$),[41, 43] obtained after the theoretical fittings of R-space experimental data (EXAFS data) of the synthesized samples is shown in Fig. 7(c). Further, $k^2\chi(R)$ vs. R plot for the EXAFS data, taken at Ti K-edge of the synthesized $TiCo_{1-x}Ni_xSb$ (x=0, 0.01, 0.02, 0.03, 0.04, and 0.06) samples are analyzed and presented in Fig. 8(a). Corresponding $k^2\chi(k)$ data are given in the SI (Fig. S8). The peaks in R-space data at around 2.25 Å and 2.85 Å are observed due to Ti-Co and Ti-Sb bonds (Fig. 8(a) ) [41]. The variation in bond lengths and $\sigma^2$ with increasing Ni concentration is depicted in Figs. 8(b) and 8(c), respectively. A scrutiny of variation in R and $\sigma^2$ (estimated from EXAFS data at both the Co and Ti K-edge) with increasing Ni concentration in $TiCo_{1-x}Ni_xSb$ (x=0, 0.01, 0.02, 0.03, 0.04, and 0.06) synthesized alloys reveals an anomaly for x=0.02. It is crucial to note that $\sigma^2$ is minimum for the $TiCo_{0.98}Ni_{0.02}Sb$ sample and increases for both the 1st (Co-Ti and Co-Sb for Co K-edge; Ti-Sb for Ti K-edge) and 2nd shells (Co-Co for Co K-edge and Ti-Co for Ti K-edge), obtained after modelling of EXAFS data at Co and Ti K-edge for synthesized $TiCo_{1-x}Ni_xSb$ (x=0, 0.01, 0.02, 0.03, 0.04, and 0.06) samples. Ni doping at the Co site of the TiCoSb causes a change in local structural parameters around Co and Ti atoms at x=0.02. A sudden change in the slope ($\frac{\Delta\sigma^2}{\Delta x}$) is observed at x=0.02. The drastic change in $\sigma^2$, estimated from the EXAFS data of the synthesized samples at Co and Ti K-edge, indicates a disordered-to-ordered local structural transition. This type of change in local structure leads to a transition from order to disorder in $VO_2$ due to temperature change [46].

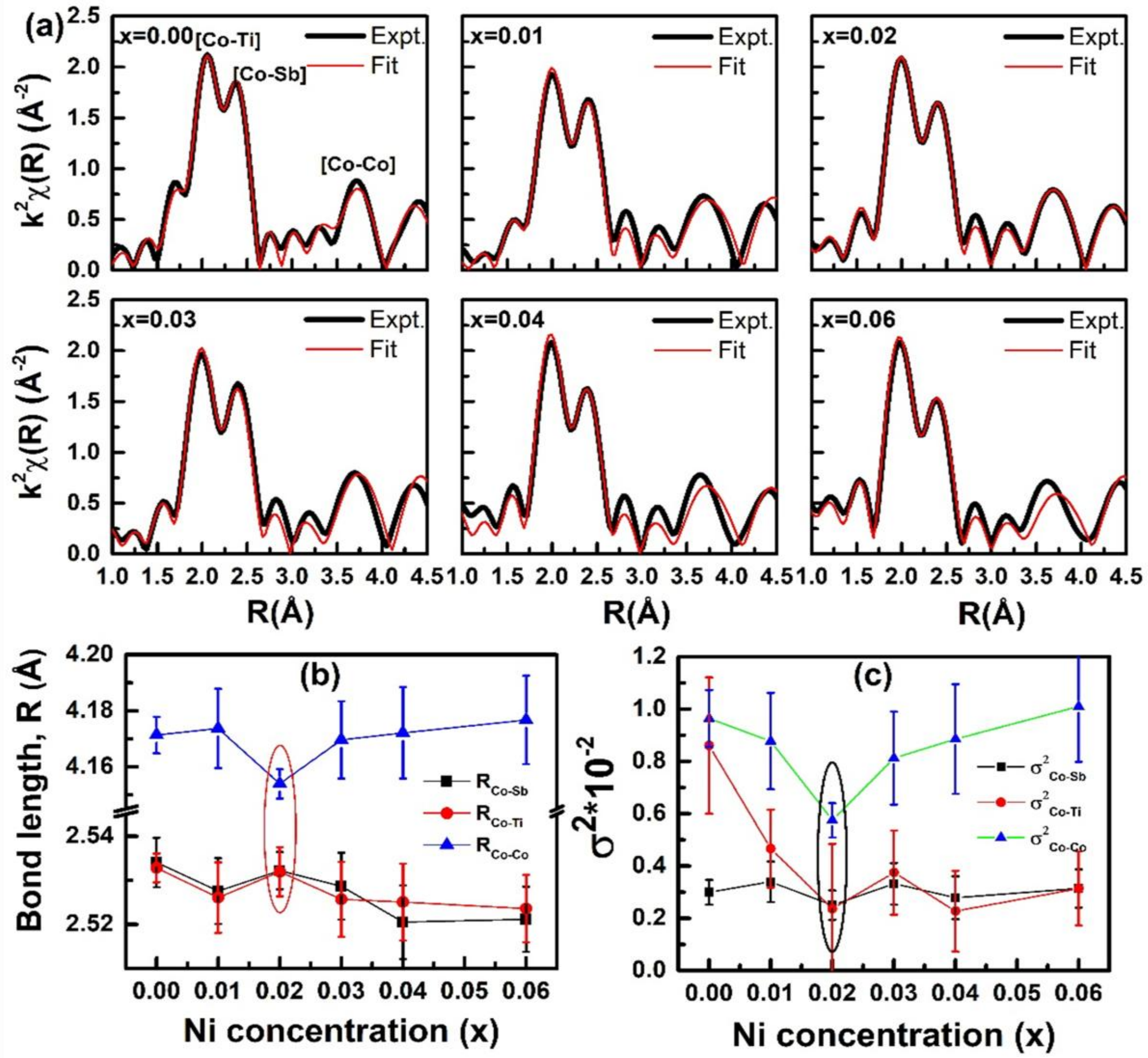

**Fig. 7. (Color online).** (a) Radial distributions, i.e., $k^2\chi(R)$ Vs. R plot of Co K-edge EXAFS data for the synthesized $TiCo_{1-x}Ni_xSb$ (x=0.0, 0.01, 0.02, 0.03, 0.04, and 0.06) HH samples. Variation of (b) Co-Ti, Co-Sb, and Co-Co bond lengths (c) corresponding disorder parameter, estimated from the fittings of EXAFS data at Co K-edge in the R-space of the synthesized $TiCo_{1-x}Ni_xSb$ HH samples.

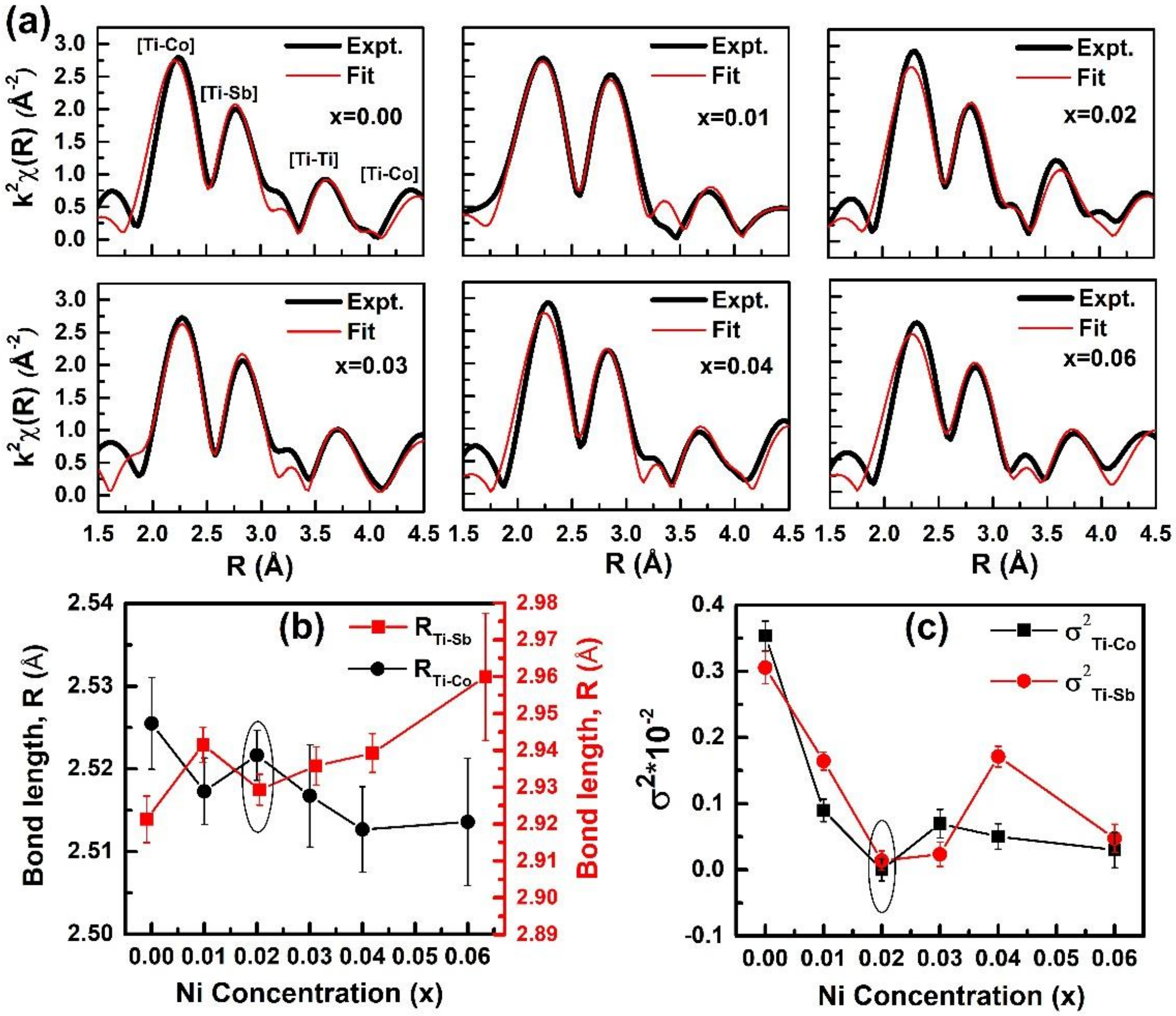

**Fig. 8. (Color online).** (a) Radial distributions, i.e., $k^2\chi(R)$ Vs. R plot of Ti K-edge EXAFS data for the synthesized $TiCo_{1-x}Ni_xSb$ (x=0.0, 0.01, 0.02, 0.03, 0.04, and 0.06) HH samples. Variation of (b) Ti-Co and Ti-Sb bond lengths (c) corresponding disorder parameter, estimated from the fittings of EXAFS data at Ti K-edge in the R-space of the synthesized $TiCo_{1-x}Ni_xSb$ HH samples.

### *4.4. Transport properties and PF*

The effect of Ni doping on transport properties of TiCoSb HH alloys is explored by the measurements of temperature-dependent resistivity (ρ(T)) and thermopower (S(T)). $\rho(T)$ of the synthesized $TiCo_{1-x}Ni_xSb$ (x=0, 0.01, 0.02, 0.03, 0.04, and 0.06) HH alloys are performed down to 10 K. Non-monotonic variation of the $\rho(T)$ data is depicted in Fig. 9(a) The change in electrical transport with temperature, as well as Ni concentration, is anazlyzed. The variation of resistivity at 150K ($\rho_{150K}$) is shown in the inset of Fig. 9(a). A careful investigation reveals that ρ(T) data undergoes a transition from metallic ($d\rho/dT > 0$) to semiconducting behaviour ($d\rho/dT < 0$) at ~150K for $0.0 \leq x \leq 0.02$. Activation energy ($E_{act}$) of the synthesized samples is estimated and presented in Fig. 9(b). The detailed procedure to calculate $E_{act}$ is provided in the SI. It is crucial to note that synthesized samples exhibit a metallic nature for the Ni concentration range $0.02 < x \leq 0.06$. The variation of $E_{act}$ indicates shifting of the Fermi surface in the synthesized samples as a function of Ni doping. A schematic to present the Fermi surface with increasing Ni concentration in $TiCo_{1-x}Ni_xSb$ HH alloys is illustrated in Fig. 9(c). However, it is crucial to note that a small amount of Ni doping in TiCoSb successfully shifts the position of the Fermi surface towards the conduction band, as obtained in the theoretical calculations in the sec. 4.1. $\rho_{150K}$, as depicted in the inset of Fig. 9(a) reveals the decrease in the value as a function of Ni concentration for $TiCo_{1-x}Ni_xSb$ synthesized samples. Increasing Ni doping concentration causes replacement of Co atoms in TiCoSb, resulting in a donation of an extra covalent electron and compensation of charge carriers, modifies the impurity band as depicted in Fig. 9(c) and position of Fermi surface as theoretically obtained in Fig. 2.

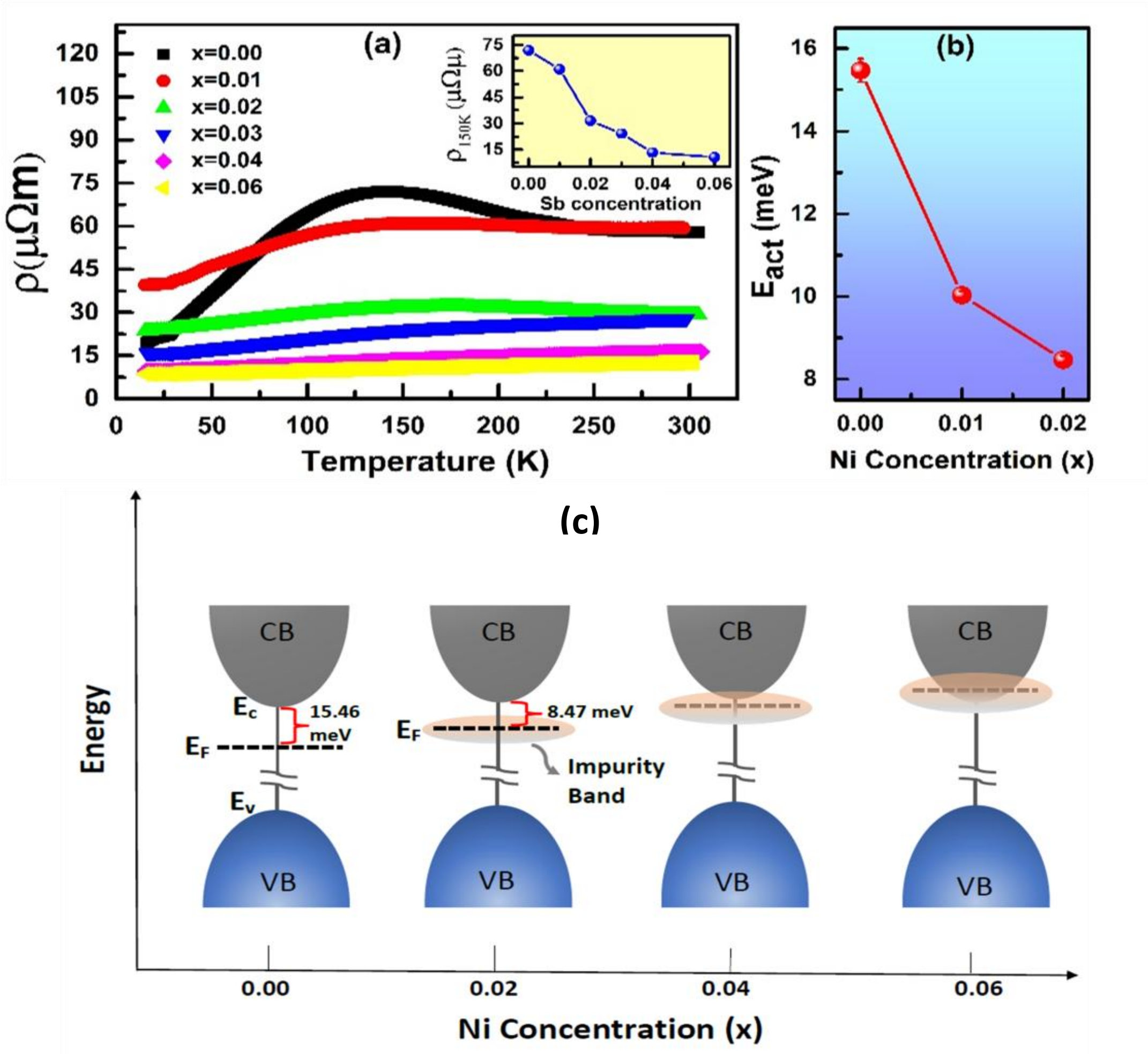

**Fig. 9. (Color online).** (a) Thermal variation of resistivity of synthesized $TiCo_{1-x}Ni_xSb$ HH samples. Inset shows the variation of resistivity at 150K temperature with Ni concentration. (b) Ni concentration dependent activation energy ($E_{act}$), estimated from the fitting of ρ(T) data at T>150K using eq. 5. (c) The energy band diagram of the synthesized $TiCo_{1-x}Ni_xSb$ HH alloy shows the formation of impurity band and the shifting of the Fermi level as charge compensation occurs.

Temperature dependent thermopower, S(T) measurements of the synthesized $TiCo_{1-x}Ni_xSb$ (x=0.0, 0.01, 0.02, 0.03, 0.04, and 0.06) samples are carried out down to 10K, employing differential technique. S(T) for the synthesized samples are depicted in Fig. 10 and the inset represents the variation of S at 300K ($S_{300K}$) as a function of Ni concentration in $TiCo_{1-x}Ni_xSb$. The negative value of S(T) for the synthesized samples indicates majority carriers involved in transport are electrons [14]. The S(T) data may also be divided into two parts, $0\leq x\leq 0.02$ and $0.02\leq x\leq 0.06$ as obtained for the other properties, mentioned above. The value of S(T) increases as a function of Ni doping till x=0.02. However, further doping causes a decrease in S(T). |S(T)| reveals that maximum and minimum values of $S_{300K}$ are achieved for x=0.02 and x=0.06 of the synthesized $TiCo_{1-x}Ni_xSb$ (x=0.0, 0.01, 0.02, 0.03, 0.04, and 0.06) HH alloys, respectively. It is crucial to note that the variation in $S_{300K}$~ 36% is observed for the sample x=0.02 with respect to the sample 0.06. The variation in wt% of the embedded phases, estimated employing Rietveld refinement, reveals that the sample with Ni doping x=0.02 contains the maximum host phase among the synthesized alloys. The steep variation in |S(T)| as a function of temperature, and decrease in value with increasing doping concentration, indicate that synthesized samples are degenerate in nature [47]. However, the carriers involved in transport properties in TiCoSb may be considered as bipolar in nature [14]. Increasing Ni doping concentration till x=0.02 on $TiCo_{1-x}Ni_xSb$ samples causes a change in conduction from bipolar to unipolar due to compensation in charge, and develops a single channel that contributes to *S* [47]. Further doping for the range $0.02\leq x\leq 0.06$ causes a large enhancement in carrier concentration, reduces the value of *S,* according to the relation $S \propto n^{-\frac{2}{3}}$ where *n*=carrier concentration [18], considering a degenerate semiconductor.[46] Xie et al. observed similar thermal variation in *S* data for $Ti_{0.5}Zr_{0.25}Hf_{0.25}Co_{1-x}Ni_xSb$ (x=0-0.05) alloys due to an increase in n [47].

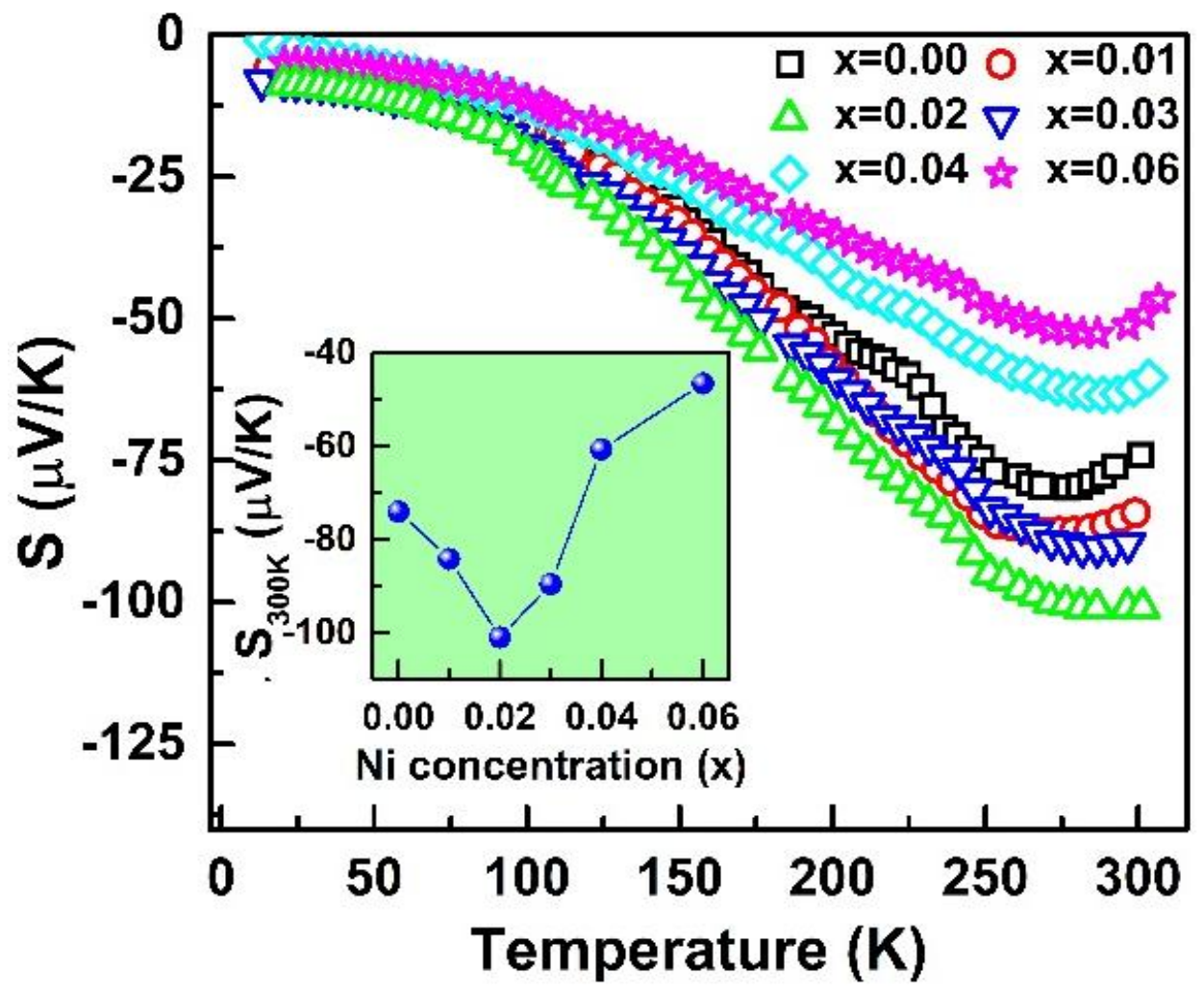

**Fig. 10. (Color online).** Thermal variation of thermopower (S(T)) of the synthesized $TiCo_{1-x}Ni_xSb$ HH alloy. Inset shows Ni concentration dependent thermopower value at 300K.

In order to reveal the in-depth characteristic of the non-monotonic ρ(T) data, the low temperature metallic and high temperature semiconducting parts are fitted using two models for $0\leq x\leq 0.02$ of the synthesized $TiCo_{1-x}Ni_xSb$ samples. The following two models are employed to fit the data, [48]

for the region $\frac{d\rho}{dT} > 0\ i.e. T < 150K$ :

$$\rho(T) = \rho_0^m + aT + bT^2 \tag{4}$$

and for the region, $\frac{d\rho}{dT} < 0\ i.e. T > 150K$:

$$\rho(T) = \rho_0^s \exp\left(\frac{-E_{act}}{2K_BT}\right). \tag{5}$$

Where $\rho_0^m$ is residual resistivity in the metallic region, and temperature independent. *a*, and *b* represent the electron-phonon (e-ph) and the electron-electron (e-e) scattering coefficients. $\rho_0^s$ is a pre-exponential term, and is determined by the material's properties. $E_{act}$ represents the activation energy for the semiconducting region. The detailed implications of the estimated $E_{act}$ and its correlation with theoretical data are explained in the preceding paragraph. However, the metallic region for $0.03\leq x\leq 0.06$ of the synthesized $TiCo_{1-x}Ni_xSb$ samples is fitted using the eq. 4.

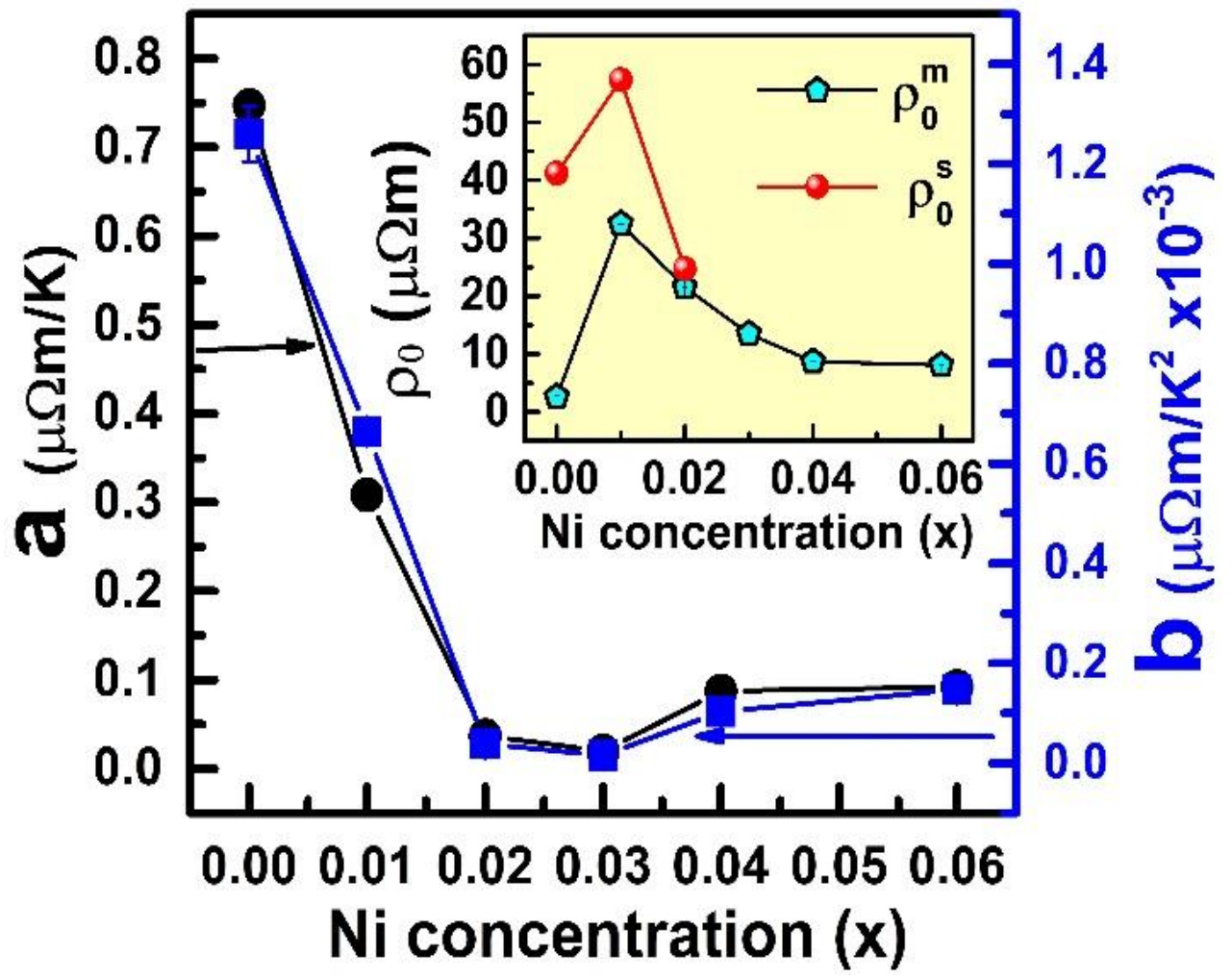

**Fig. 11. (Color online).** Ni concentration dependent electron-phonon, 'a' and electron-electron, 'b' scattering coefficients estimated from the fitting of ρ(T) data using eq. 4. Inset shows the variation of residual resistivity, $\rho_0^m$ and $\rho_0^s$ in the metallic and semiconducting region estimated from the fitting of ρ(T) data using eqs. 4 and 5, respectively.

The estimated values of *a* and *b* coefficients for all the metallic regions as a function of Ni concentration are plotted in Fig. 11. Both the *a* and *b* values decrease with increasing Ni doping till x=0.02 for the synthesized samples and almost saturates for further increases in Ni concentration. A decrease in *a* value signifies a reduced e-ph interaction and results in an increase of carrier mobility [48]. Metallicity increases with increasing Ni concentration in $TiCo_{1-x}Ni_xSb$ synthesized alloy, as reflected by a decrease in e-e scattering coefficient, *b*. The values of $\rho_0^m$ and $\rho_0^s$, obtained by fitting the experimental $\rho(T)$ data of the synthesized samples employing eq. 4 and eq. 5, are depicted in the inset of Fig. 11 as a function of Ni concentration. $\rho_0^m$ is related to crystal imperfection and grain boundary scattering. The maximum value of $\rho_0^m$ is obtained for x=0.01 sample and decreases on either side with increasing Ni concentration in synthesized $TiCo_{1-x}Ni_xSb$ HH alloys. The value of resistivity at low temperature (~20K) perfectly follows the trend of variation in $\rho_0^m$ as a function of Ni concentration. $\rho_0^s$ is obtained for three semiconducting samples for T>150K. The highest value is obtained for x=0.01 and a sudden drop is observed for x=0.02 of the $TiCo_{1-x}Ni_xSb$ samples. A decrease in $\rho_0^s$ value for x=0.02 is correlated to the effective density of states and effective mass of a degenerate semiconductor [49]. However, $\rho$ in a degenerate semiconductor is the result of a carrier conduction in a complex interplay of band structure and carrier concentration

Where $k_B$ is the Boltzmann constant and η represents the reduced Fermi energy, which is derived from the measured S(T) using the following equation,

$$S = \pm\frac{k_B}{e}\left(\frac{\left(r+\frac{5}{2}\right)F_{r+3/2}(\eta)}{\left(r+\frac{3}{2}\right)F_{r+1/2}(\eta)} - \eta\right) \quad (7)$$

Where $F_n(\eta)$ is the nth order the Fermi integral,

$$F_n(\eta) = \int_0^\infty \frac{x^n}{1+e^{x-\eta}}dx, \eta = \frac{E_f}{k_B T} \quad (8)$$

*L(T)* is depicted in Fig. 12(a), for different Ni concentrations in synthesized $TiCo_{1-x}Ni_xSb$. The variation of L(T) at high and low temperatures is presented in Figs. 12(b) and 12(c). *L(T)* increases with increasing temperature due to the heightened value of phonon scattering [52]. It is crucial to mention that there is a sharp change in the slope of the L(T) data at ~125K ($L_{125K}$). The anomalous behavior and change in the slope may be related to the change of structural order or temperature-dependent phase transition in the synthesized alloys [52]. The in the presence of different types of scattering. In order to reveal the effect of scattering mechanism, impurities, and band structure on the transport properties of the synthesized $TiCo_{1-x}Ni_xSb$ alloys, temperature-dependent Lorentz number (*L(T)*) is estimated from the temperature dependent *S(T)*, using the equation, [50, 51]

$$L = \left(\frac{k_B}{e}\right)^2\left(\frac{\left(r+\frac{7}{2}\right)F_{r+5/2}(\eta)}{\left(r+\frac{3}{2}\right)F_{r+1/2}(\eta)} - \left[\frac{\left(r+\frac{5}{2}\right)F_{r+3/2}(\eta)}{\left(r+\frac{3}{2}\right)F_{r+1/2}(\eta)}\right]^2\right). \quad (6)$$

value of *L(T)* of $TiCo_{1-x}Ni_xSb$ (x=0, 0.01, 0.02, 0.03, 0.04 and 0.06), increases in the range $0.0\leq x\leq 0.02$ and decreases for $0.02\leq x\leq 0.06$, below 125K (Fig. 12(b)). But, opposite behavior is observed above 125K, where L(T) is reduced for $0.0\leq x\leq 0.02$ and enhanced in the range $0.0\leq x\leq 0.06$ (Fig. 12(c)). However, the variation in *L(T)* as a function of Ni doping at the transition temperature (Fig. 12(d)) reveals that the variation is similar to that obtained at high temperature, T>125K. The *L(T)* value at room temperature is comparable to the Sommerfeld value $L_0 \sim 2.44 \times 10^{-8}\, V^2K^{-2}$ [52, 53]. Carrier concentration of semiconducting samples, for Ni doping range $0.0 \leq x \leq 0.02$ in $TiCo_{1-x}Ni_xSb$ synthesized samples, increases with temperature. Metallicity in the synthesized degenerate semiconducting samples is also increased due to the compensation to overcompensation of charge carriers with Ni doping, causing an enhancement in carrier concentration. L(T) tends to $L_0$ with increasing temperature for T>125 K, and L(T)~$L_0$ is obtained at room temperature for the synthesized samples. However, the variation in L(T) as a function of Ni concentration at high temperature, Fig. 12(c), is corroborated with the presence of embedded phases (Fig. 5(b)). Higher value of L is also observed in gold samples due to the presence of impurities [52, 54].

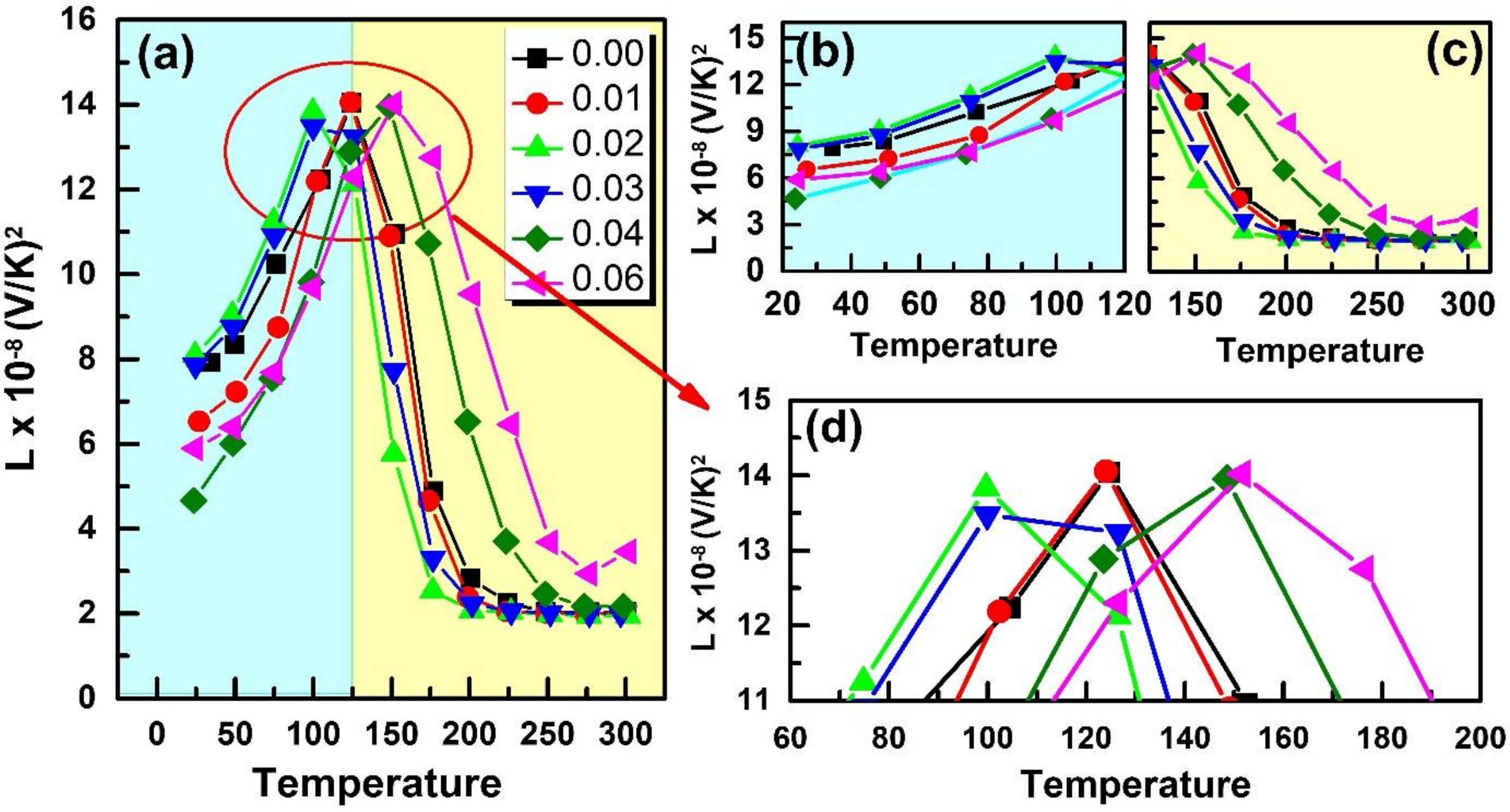

**Fig. 12. (Color online).** (a) Thermal variation of Lorentz number (L) for the synthesized $TiCo_{1-x}Ni_xSb$ (x=0, 0.01, 0.02, 0.03, 0.04, and 0.06) polycrystalline HH alloys. Enlarged view of L for (b) T<125 K and (c) T>125 K. (d) shows the enlarged view of L(T) near transition temperature, i.e., T~125K.

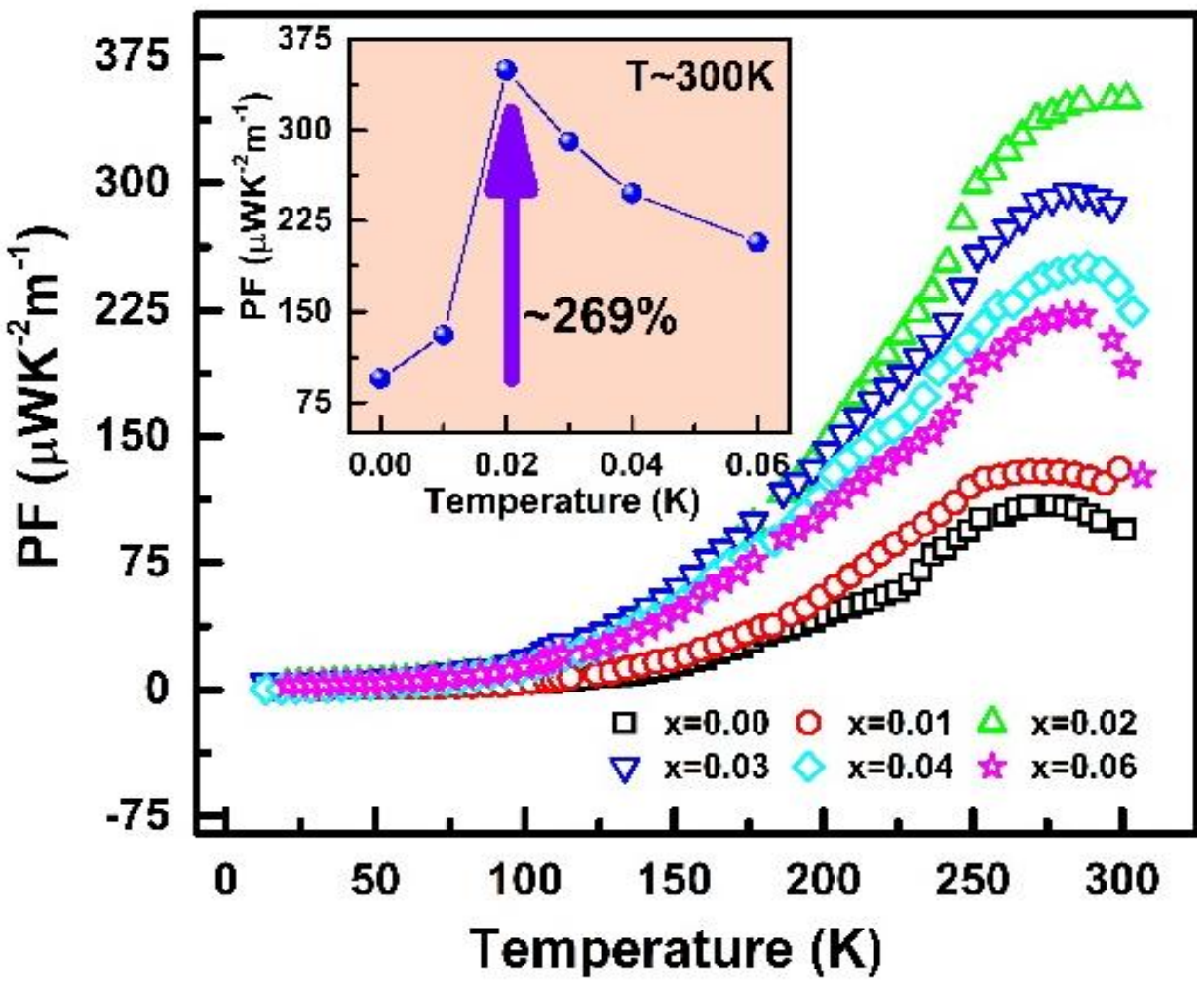

**Fig. 13. (Color online).** Thermal variation of power factor (PF) of the synthesized $TiCo_{1-x}Ni_xSb$ HH alloy. Inset shows the variation of PF with Ni concentration at room temperature (300K).

Replacement of Co atom by Ni atom in TiCoSb causes a change in the nature of ρ(T) behaviour, semiconducing to metallic. The embedded CoTi phases in the $TiCo_{1-x}Ni_xSb$ matrix have a significant impact on the transport properties of the synthesized samples [36]. The metallic behaviour in the ρ(T) graph for $0.0 \leq x \leq 0.02$ below 150 K may be related to the presence of metallic CoTi embedded phases, and activated behaviour triggered after 150K [36]. Theoretical calculation reveals that a minute amount of Ni doping causes a drastic shift in the Fermi surface and supports the ρ(T) behaviour as a function of Ni concentration. The estimated $\rho_0^m$, from experimental ρ(T) data, also indicates that there is a drastic change in band structure and density of state for x=0.02 of synthesized $TiCo_{1-x}Ni_xSb$ sample. Romaka et al. reported a transition from semiconductor to metallic behaviour in ρ(T) data of the $TiCo_{1-x}Ni_xSb$ system for $x \geq 0.1$ [15]. A similar type of transition is also identified for TiNiSn and ZrNiSn HH alloys owing to the substitution of Ti/Zr with Sc [55]. However, Han et al. reported that structural order influences the metal-to-insulator transition in $VO_2$ [46]. The EXAFS data confirms the disordered-to-ordered transition in $TiCo_{1-x}Ni_xSb$ synthesized alloy for x=0.02. This transition causes a drastic change in the DOS, band structure, e-e, and e-ph interaction. The change in electronic structure is evident from the estimated parameters from ρ(T). However, the anomalous behaviour in L(T) data for x=0.02 also confirms the structural transition [52].

PF ($S^2\sigma$) of $TiCo_{1-x}Ni_xSb$ (x=0, 0.01, 0.02, 0.03, 0.04, and 0.06) synthesized samples are estimated from S(T) and ρ(T) data. The non-monotonic thermal variation of PF of the synthesized samples is depicted in Fig. 13. PF follows the similar trend as observed for other experimental data, i.e., PF increases for the range $0.0 \leq x \leq 0.02$ and decreases thereafter until x=0.06. PF at room temperature ($PF_{300K}$) as a function of Ni concentration is presented in the inset of Fig. 13. $PF_{300K}$ increases from 95 $\mu WK^{-2}m^{-1}$ for x=0.0 to 350 $\mu WK^{-2}m^{-1}$ for x=0.02.

## 5. Conclusion

Theoretical and experimental investigations are performed to reveal the effect of Ni concentration on $TiCo_{1-x}Ni_xSb$ HH alloy. Long-range structural characterization is performed employing Rietveld refinement of the XRD data. The lattice parameter of TiCoSb, optimized theoretically and estimated by refinement are comparable. It is crucial to mention that structural parameters, viz., *a*, ε, relative dislocation density, *V*, wt%, and $B_{iso}$, show anomalies for $TiCo_{0.98}Ni_{0.02}Sb$ synthesized samples. The maximum host phase along with a slope-change in Ni concentration dependent lattice volume at the x=0.02 indicates a structural rearrangement. The minimum value of $B_{iso}$ also represents the least distortion for x=0.02. In-depth local structure is explored by XAS measurements at Co and Ti K-edge. The detailed analysis of XANES and EXAFS data reveals that a local structural transition from disorder to most order structure is occurred for the x=0.02 Ni concentration.

The structural order influences the transport properties as well as the electronic structure and scattering processes involved in the carrier transport. However, extra electrons are added to the system due to the replacement of Co by Ni atoms, causing a change from multi-channel to single channel conduction and an increase in carrier concentration. Theoretical investigation also supports the results as observed change in semiconducting to metallic behavior of ρ(T) data. The estimated parameters from ρ(T) data i.e., e-e and e-ph scattering coefficients, also stipulate a drastic change in DOS and electronic structure of the synthesized material at x=0.02.

The theoretical investigation shows that a minute amount of Ni doping in TiCoSb HH alloy drastically shifts the Fermi surface towards conduction bands. The impurity scattering involved in S(T) and ρ(T) data is observed on thermal variation of *L,* and a minimum is reported for $TiCo_{0.98}Ni_{0.02}Sb$, corroborating the maximum host phase as obtained from Rietveld refinement data. However, the drastic change in L(T) near about 120K may be related to the temperature-dependent structural transition in TiCoSb HH alloy. The nonmonotonic S(T) of the synthesized samples also indicates the anomalous behavior for 2% Ni doping in TiCoSb HH alloy. The anomaly as observed in the structural, transport, and estimated parameters, supports the XAS result, i.e., an isoelectronic transition from disordered to ordered structure in the local environment for x=0.02. Further, a simultaneous optimization, i.e., increases in S and decreases in ρ, concomitantly enhance the PF~269% for the synthesized $TiCo_{0.98}Ni_{0.02}Sb$ sample.

## Acknowledgement

This work is supported by the Science and Engineering Research Board (SERB) (File Number: EEQ/2018/001224) ndia in the form of sanctioning research project. The financial support of UGC-DAE-CSR Kalpakkam, India (Ref: CRS/2021-22/04/639, CRS/2022-23/04/893) is acknowledged in the form of research project grants. The authors gratefully acknowledge Dr. Prasanjit Samal, NISER Bhubaneswar, India, for providing access to the VASP software used in this study. We are thankful to Dr. Aritra Banerjee, University of Calcutta, for the discussion and help in carrying out resistivity and thermopower measurements. We gratefully acknowledge late Avijit Jana for his support in carrying out the experiments and engaging in valuable discussions. We remember his contributions with deep respect and sorrow. Further, we would also like to acknowledge the synchrotron beamline, BL-9, INDUS-2, Raja Ramanna Centre for Advanced Technology, for providing the facility of XAS measurement.

---

Corresponding author at: *Vidyasagar Metropolitan College, Kolkata-700006, India.*

Corresponding author E-mail address: kartick.phy09@gmail.com (Kartick Malik)

# Supplementary Information

## The influence of Ni doping on local structure and electronic properties of TiCoSb: Effect on transport properties

Suman Mahakal[1], Pallabi Sardar[1, 2], Diptasikha Das[2], Subrata Jana[3], Swapnava Mukherjee[1, 2], Biplab Ghosh[4], Shamima Hussain[5], Santanu K. Maiti[6], and Kartick Malik*[1]

[1] *Department of Physics, Vidyasagar Metropolitan College, Kolkata-700006, India.*

[2]*Department of Physics, ADAMAS University, Kolkata-700126, India.*

[3]*Institute of Physics, Faculty of Physics, Astronomy and Informatics, Nicolaus Copernicus University, Torun, ul. Grudziadzka 5, 87-100 Torun, Poland.*

[4]*Beamline Development & Application Section, Bhabha Atomic Research Centre, Trombay, Mumbai-400 085, India*

[5] *UGC-DAE Consortium for Scientific Research, Kalpakkam Node, Kokilamedu 603 104, Tamil Nadu, India.*

[6] *Physics and Applied Mathematics Unit, Indian Statistical Institute, 203 Barrackpore Trunk Road, Kolkata-700 108, India.*

* Corresponding authors

Kartick Malik - Email: kartick.phy09@gmail.com

# Table of Contents

## 1. Estimation of Dislocation density ($N_D$), using modified Williamson-Hall plot

Relative dislocation density of the $TiCo_{1-x}Ni_xSb$ (x=0.0, 0.01, 0.02, 0.03, 0.04, and 0.06) synthesized samples are estimated from the X-ray diffraction (XRD) data. The Dislocation density ($N_D$) of the synthesized samples are obtained using the following equation [1, 2]:

$$\Delta K = \frac{0.9}{d} + \frac{\pi A^2 B_D^2}{2} N_D^{1/2} K^2 C \pm O(K^4 C^2) \tag{S1}$$

$C$ = Average value of dislocation contrast factor = $C_{h00} \dfrac{1-q(h^2k^2+k^2l^2+h^2l^2)}{(h^2+k^2+l^2)^2}$.

*h, k, l* ➔ Millar indices of crystal plane corresponding to the XRD peak.

$C_{h00}$ ➔ Average value of dislocation contrast factor corresponding to the *h00* reflection plane.

*Q* ➔ Constant parameter

*A* ➔ Parameter associated to effective outer cut-off radius of crystal dislocations

$B_D$ ➔ Burgers vector

*d* ➔ Average crystallite size

It is assumed that the parameters q, *A* and $B_D$ do not change significantly for samples $TiCo_{1-x}Ni_xSb$ (x=0, 0.01, 0.02, 0.03, 0.04, 0.06) due to minute amount of doping. Here, we have estimated relative dislocation densities $N_D/N_{D|x=0}$ to avoid the unknown constant terms.

$$\left(\frac{N_D}{N_{D|=0}}\right) = \left(\frac{m_x}{m_{x=0}}\right)^2.$$

Where $m_x$ is the slope of the linear fittings of '*ΔK* versus $K^2C$' data, obtained for $TiCo_{1-x}Ni_xSb$ (x=0.0, 0.01, 0.02, 0.3, 0.04, and 0.06) samples. And $m_{x=0}$ denotes the slope of linear fittings of '*ΔK* versus $K^2C$' data for x = 0.

***Figure S1***. *Modified Williamson-Hall plots and corresponding fittings for the polycrystalline* $TiCo_{1-x}Ni_xSb$ *(x=0.0, 0.01, 0.02, 0.03, 0.04, and 0.06) half-Heusler alloys. Solid line represents best linear fit.*

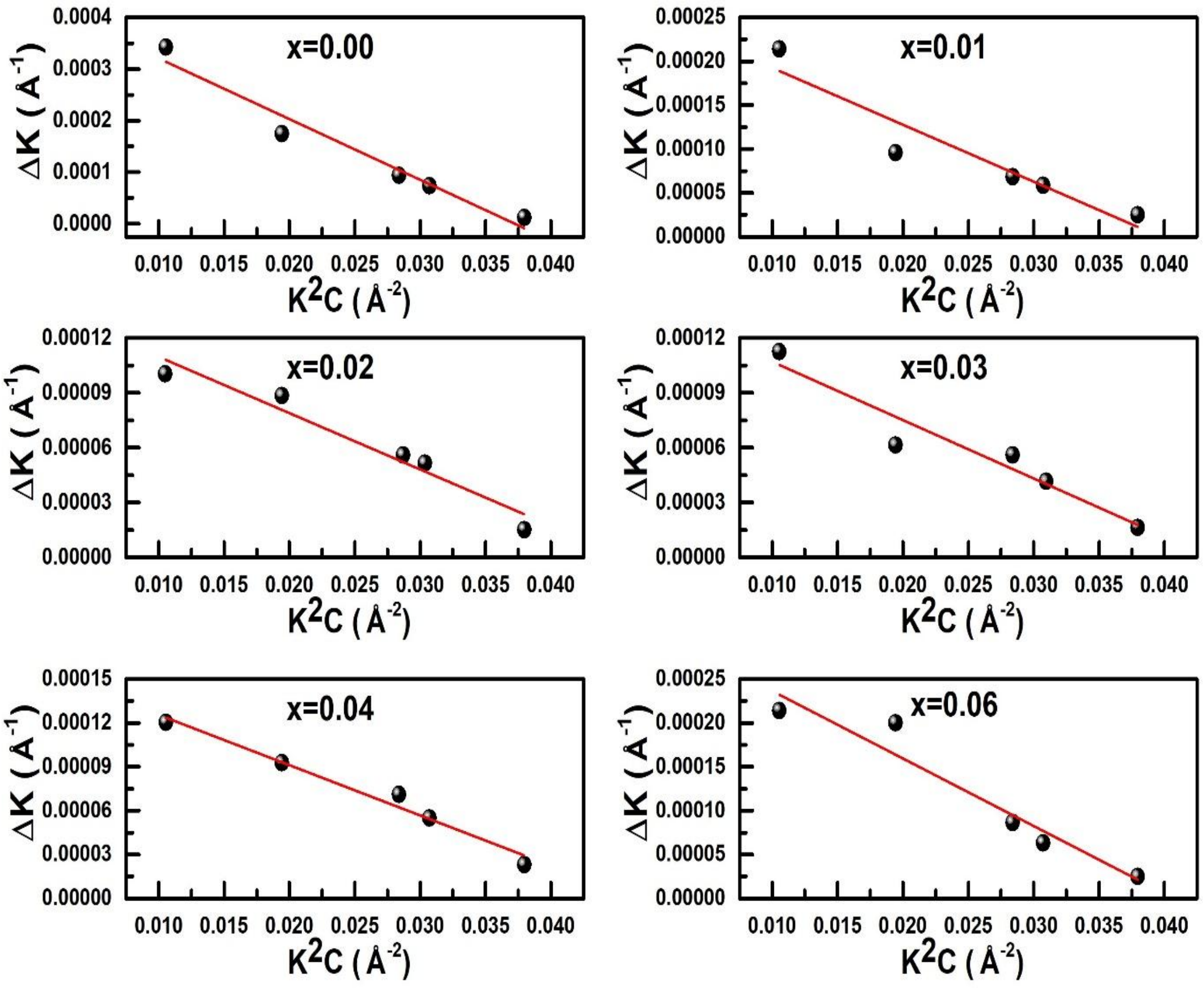

## 2. Structural parameters and fitted X-ray diffraction data, obtained after Rietveld refinement

***Figure S2.*** *X-ray diffraction patterns of $TiCo_{1-x}Ni_xSb$ (x = 0.00, 0.01, 0.02, 0.03, 0.04, and 0.06) synthesized samples at room temperature, fitted employing Rietveld refinement using FullProf software and the corresponding refinement parameters.*

(a)

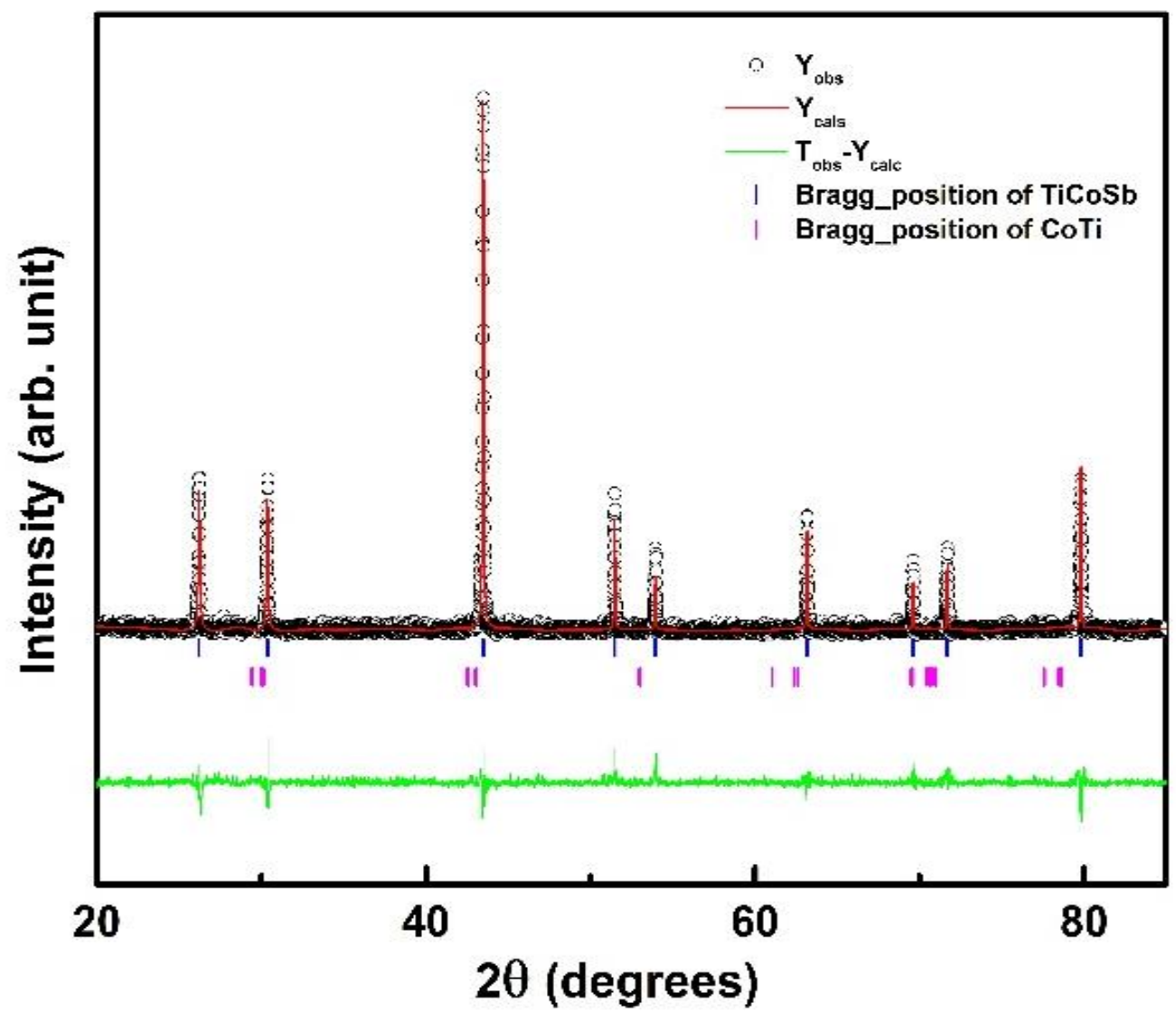

| | RT |
|---|---|
| Phase | TiCoSb |
| Cell (Å) | *a* =b= *c*: 5.8832 |
| *Site occupancy: Ti/Co/Sb* | 1/1/1 |
| $B_{iso}$ (Å$^2$) | 0.675 |
| *TiCoSb Phase (%)* | 99.44 |
| *TiCo Phase (%)* | 0.54 |
| $R_p$ (%) | 28.69 |
| $R_{wp}$(%) | 22.13 |
| $R_{exp}$(%) | 13.17 |
| Gof | 2.83 |

(b)

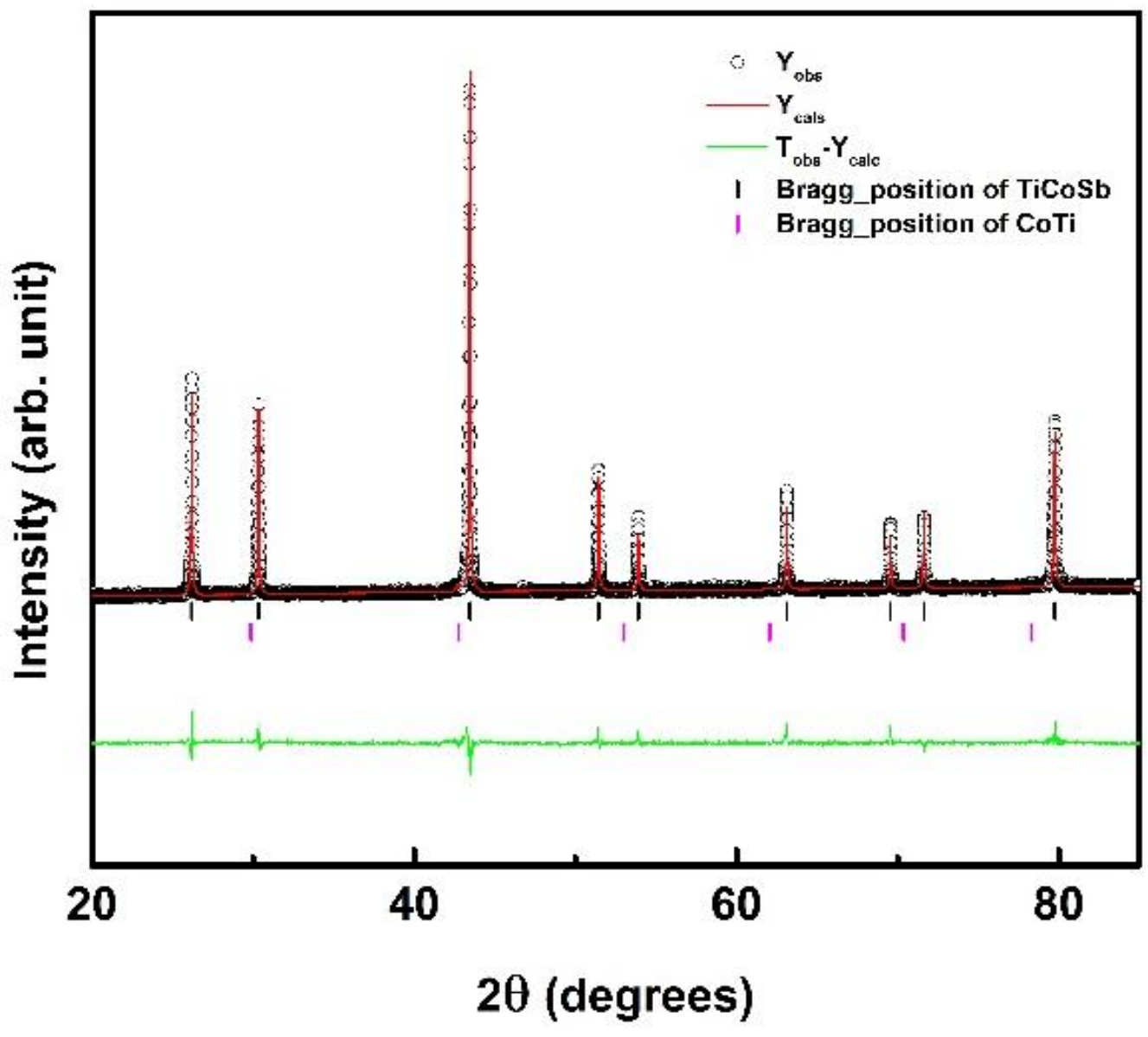

| | RT |
|---|---|
| Phase | $TiCo_{0.97}Ni_{0.03}Sb$ |
| Cell (Å) | *a* =b= *c*: 5.88279 |
| *Site occupancy: Ti/Co/Sb* | 1/0.97/0.03/1 |
| $B_{iso}$ (Å$^2$) | 0.69 |
| *TiCoSb Phase (%)* | 99.77 |
| *TiCo Phase (%)* | 0.23 |
| $R_p$ (%) | 38.8 |
| $R_{wp}$(%) | 21.8 |
| $R_{exp}$(%) | 11.84 |
| Gof | 3.39 |

(c)

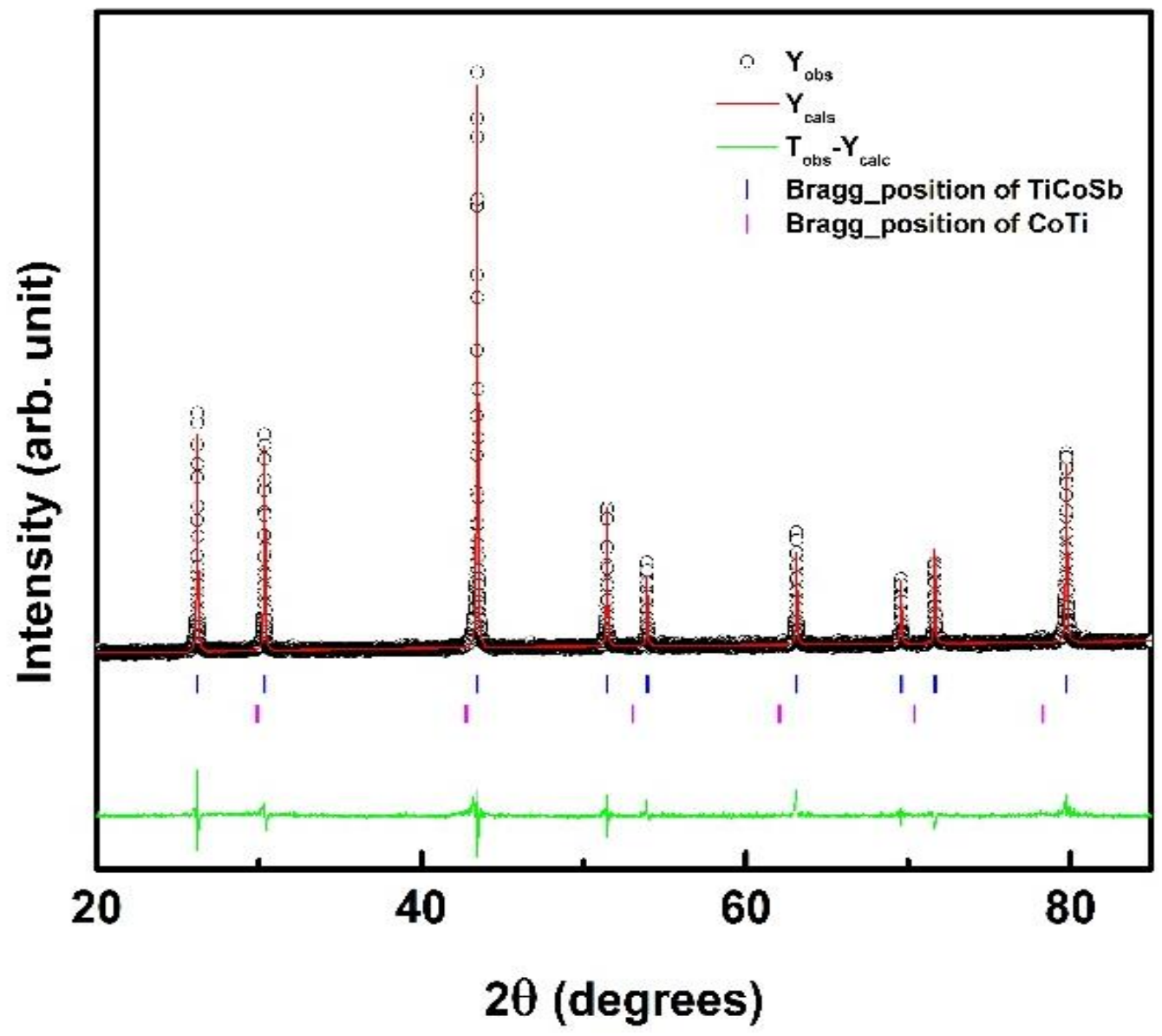

| | **RT** |
|---|---|
| Phase | $TiCo_{0.98}Ni_{0.02}Sb$ |
| Cell (Å) | *a* =b= *c*: 5.8829 |
| *Site occupancy: Ti/Co/Sb* | 1/0.98/0.02/1 |
| $B_{iso}$ (Å$^2$) | 0.67 |
| *TiCoSb Phase (%)* | 99.85 |
| *TiCo Phase (%)* | 0.15 |
| $R_p$ (%) | 35.9 |
| $R_{wp}$(%) | 22.7 |
| $R_{exp}$(%) | 11.92 |
| Gof | 3.62 |

(d)

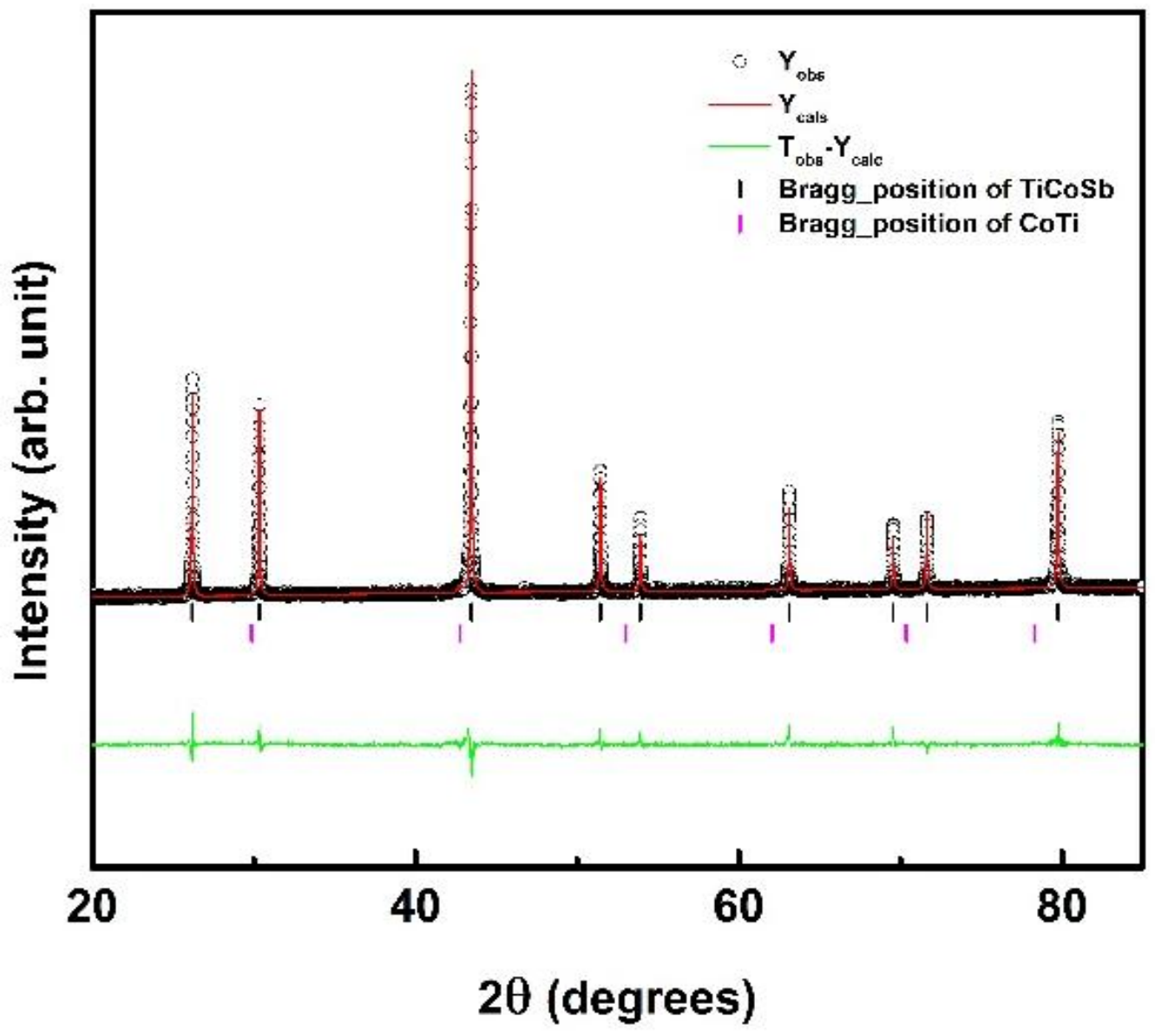

| | **RT** |
|---|---|
| Phase | $TiCo_{0.97}Ni_{0.03}Sb$ |
| Cell (Å) | *a* =b= *c*: 5.88279 |
| *Site occupancy: Ti/Co/Sb* | 1/0.97/0.03/1 |
| $B_{iso}$ (Å$^2$) | 0.69 |
| *TiCoSb Phase (%)* | 99.77 |
| *TiCo Phase (%)* | 0.23 |
| $R_p$ (%) | 38.8 |
| $R_{wp}$(%) | 21.8 |
| $R_{exp}$(%) | 11.84 |
| Gof | 3.39 |

(e)

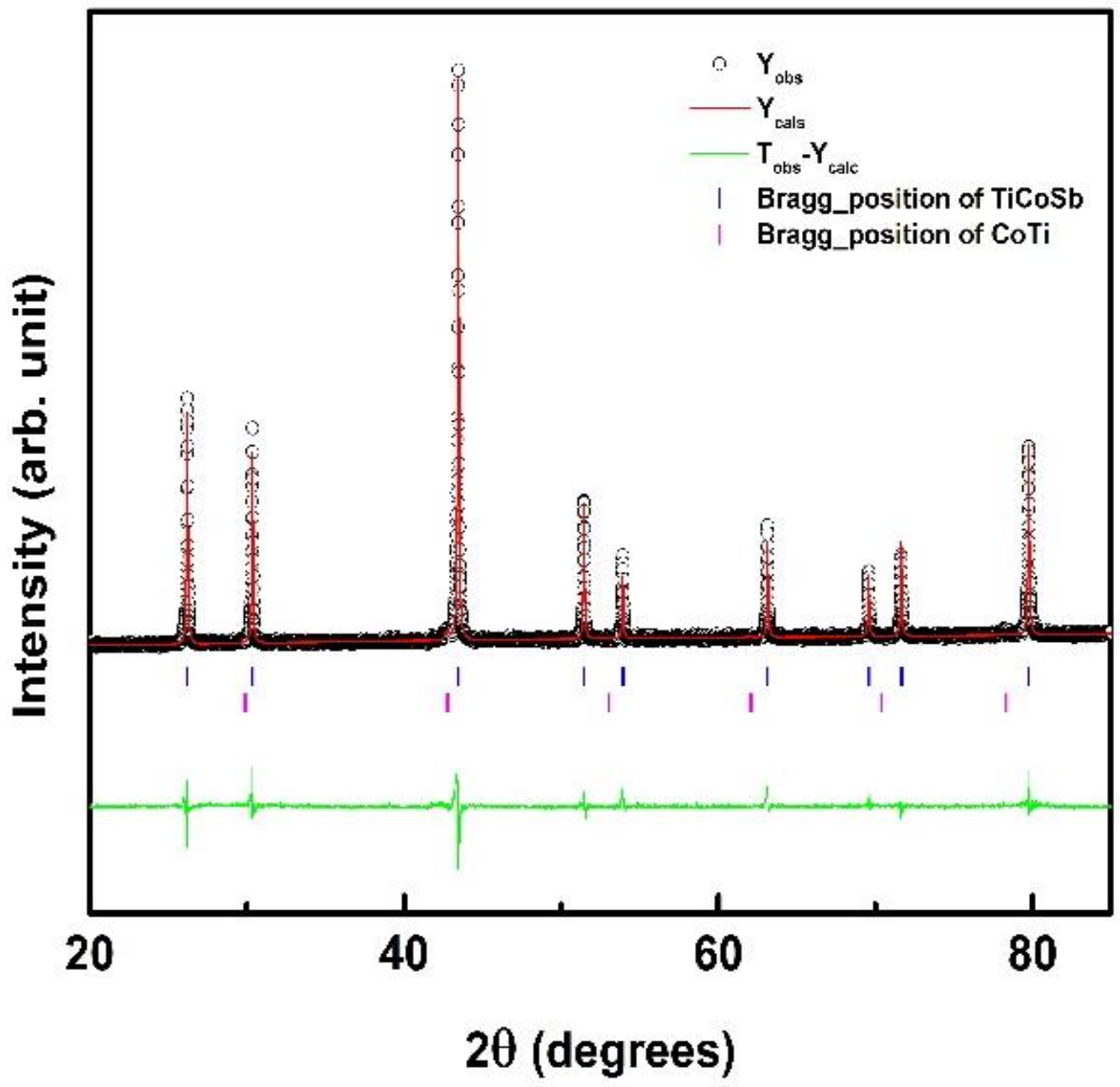

| | RT |
|---|---|
| Phase | $TiCo_{0.96}Ni_{0.04}Sb$ |
| Cell (Å) | *a* =b= *c*: 5.88273 |
| *Site occupancy: Ti/Co/Sb* | 1/0.96/0.04/1 |
| $B_{iso}$ (Å$^2$) | 0.69 |
| *TiCoSb Phase (%)* | 99.58 |
| *TiCo Phase (%)* | 0.42 |
| $R_p$ (%) | 31.7 |
| $R_{wp}$(%) | 21.2 |
| $R_{exp}$(%) | 10.83 |
| Gof | 3.82 |

(f)

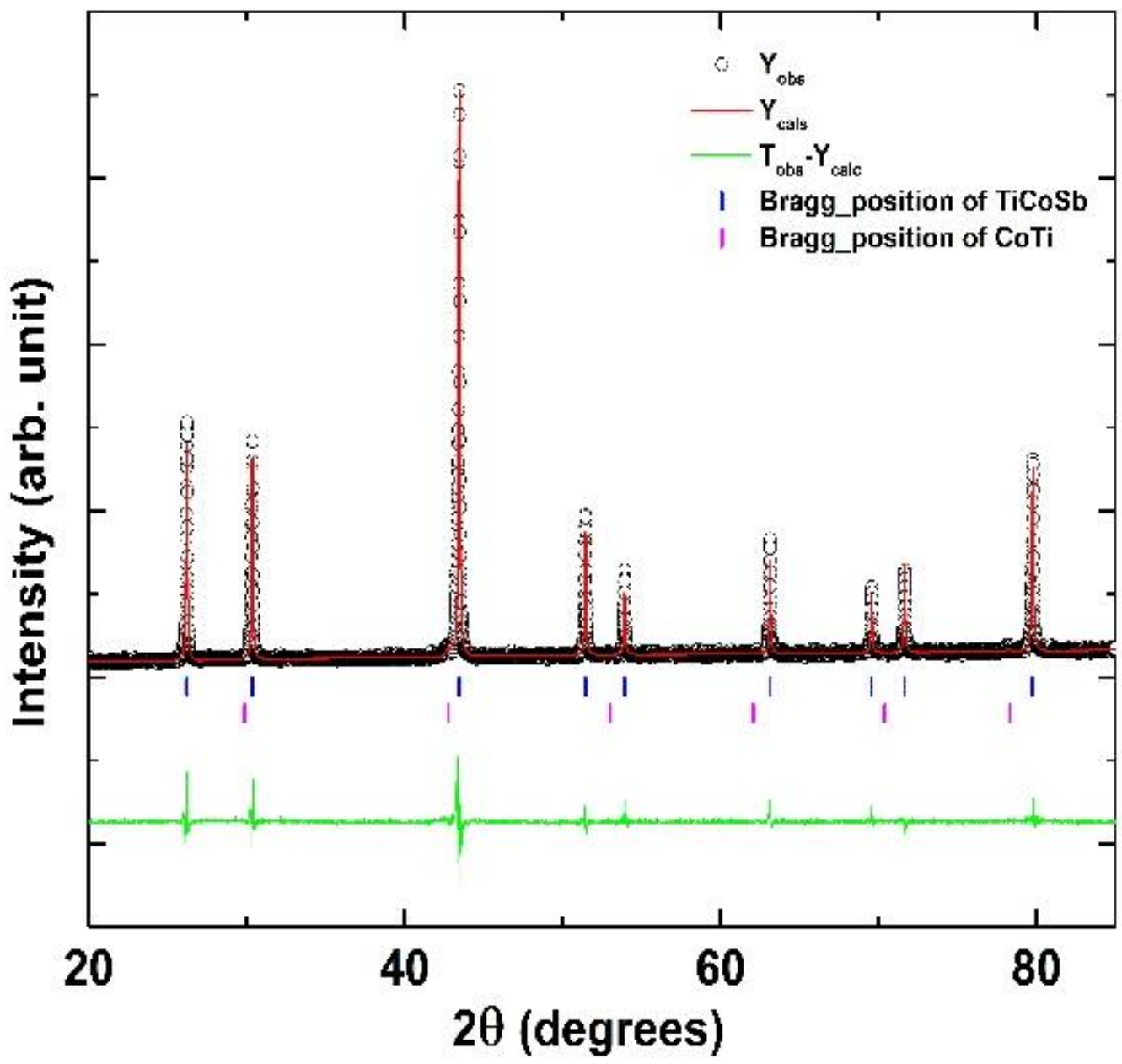

| | RT |
|---|---|
| Phase | $TiCo_{0.94}Ni_{0.06}Sb$ |
| Cell (Å) | *a* =b= *c*: 5.88270 |
| *Site occupancy: Ti/Co/Ni/Sb* | 1/0.94/0.06/1 |
| $B_{iso}$ (Å$^2$) | 0.71 |
| *TiCoSb Phase (%)* | 99.54 |
| *TiCo Phase (%)* | 0.46 |
| $R_p$ (%) | 30.8 |
| $R_{wp}$(%) | 22.8 |
| $R_{exp}$(%) | 10.15 |
| Gof | 3.68 |

## 3. Williamson-Hall Plots of the X-ray diffraction data

Williamson–Hall (WH) analysis was employed on X-ray diffraction (XRD) data to estimate the crystalline strain (ε) and average crystallite size (D). The Williamson–Hall equation is given by [3],

$$\beta cos\theta = \frac{k_B\lambda}{D} + 4\varepsilon\, sin\theta. \tag{S2}$$

β, $k_B$ and λ are the broadening in diffraction peak of the synthesized sample, the Boltzmann constant, and the wavelength of X-ray, respectively. The crystallite size (D) and microstrain (ε) were determined from the linear fit of βcosθ versus 4sinθ (Williamson–Hall plot) plot.

***Figure S3.*** *Williamson-Hall plots of the X-ray diffraction data for synthesized $TiCo_{1-x}Ni_xSb$ (x=0, 0.01, 0.02, 0.03, 0.04, and 0.06) polycrystalline half-Heusler alloys. Fitting qualities indicate a good fit for x=0.02 sample.*

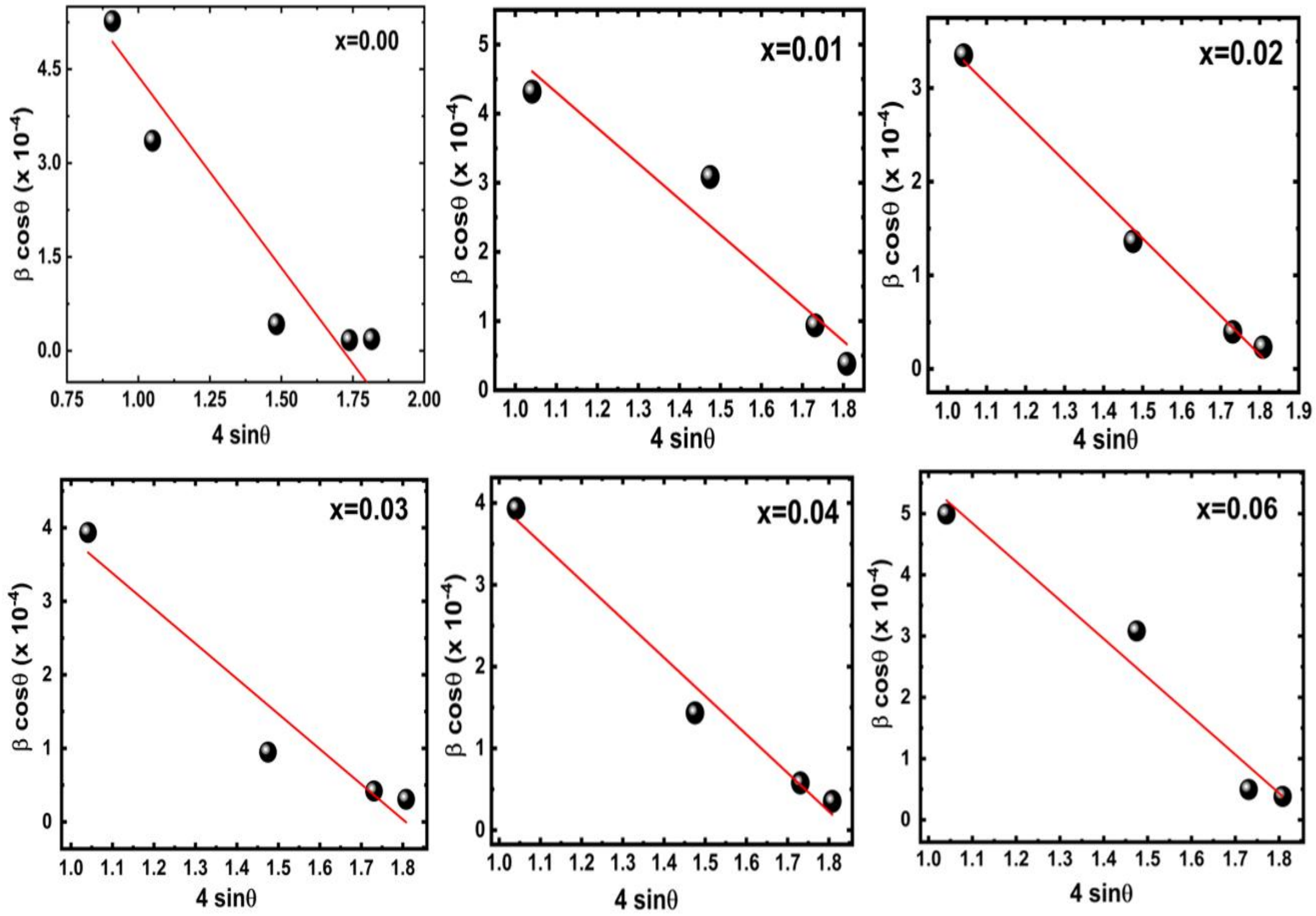

## 4. Detailed analysis of X-ray absorption spectroscopy (XAS) data for the synthesized $TiCo_{1-x}Ni_xSb$ (x=0, 0.01, 0.02, 0.03, 0.04, and 0.06) samples at Ti and Co K-edge

*XAS: Co K-edge*

The Co K-edge absorption occurs when X-rays have enough energy to remove a 1s electron from the Co atom. The electronic configuration of Co in TiCoSb is [Ar] $3d^9 4s^0$. Two types of processes may take place: (i) the 1s electron is excited to empty 4p states, which creates the Co K-edge feature in the X-ray absorption near edge structure (XANES) region, and (ii) electrons transition to the higher energy continuum states, producing the extended X-ray absorption fine structure (EXAFS) spectroscopy signal, reflecting the local atomic arrangements. Although there are unfilled 3d and 4s states below 4p, Quantum selection rules forbid 1s → 3d and 1s → 4s transitions in spite of the presence of 3d and 4s electronic states. Only the 1s → 4p transition is allowed, which is responsible for the observed K-edge absorption spectra.

***Figure S4.*** *(a) Electronic configuration of (a) Co and (b) Ti in TiCoSb HH alloy. All possible transitions from 1s to higher excitation states are highlighted by black, blue, and red arrows. The transitions, highlighted by the red arrow, are allowed according to the quantum selection rule.*

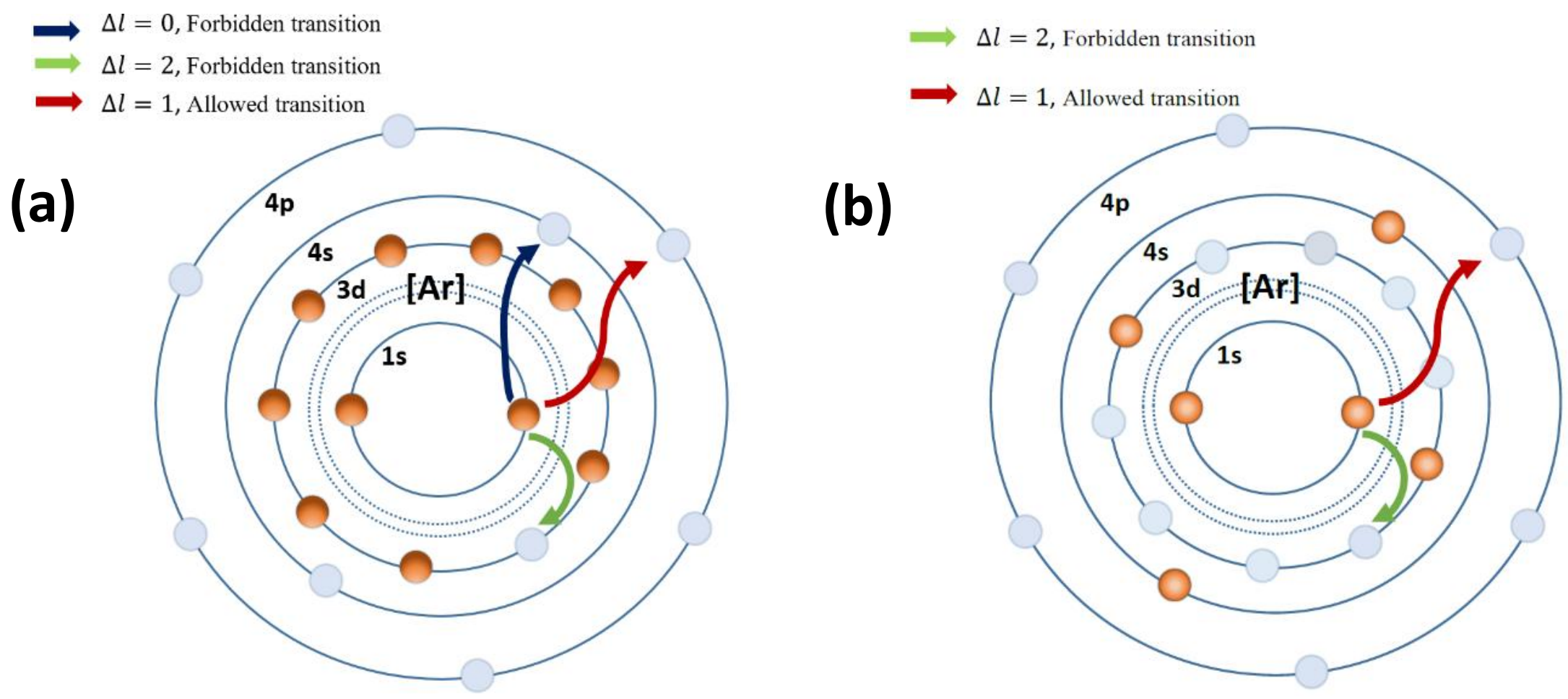

*XAS: Ti K-edge*

The Ti K-edge spectrum of the TiCoSb HH alloys is a result of the excitation of the 1s electrons to the unoccupied 4p levels as the final state. The empty 3d orbitals lying below the 4p level could provide required holes for the excitation of core electrons. However, 1s→3d (s→d, $\Delta l = 2$) transition is forbidden owing to the quantum selection rules, i.e., only $\Delta l = \pm 1$ transition is allowed. Therefore, the 1s→4p transition is the only possible route by which the 1s electron could be excited to the higher lying levels.

***Figure S5.*** *X-ray absorption fine structure (XAFS) spectroscopy, showing the absorption coefficient (μ) as a function of photon energy (E) of the* $TiCo_{1-x}Ni_xSb$ *(x=0, 0.01, 0.02, 0.03, 0.04, and 0.06) polycrystalline HH samples at Co K-edge, indicating the X-ray absorption near edge structure (XANES) spectroscopy and extended X-ray absorption fine structure (EXAFS) spectroscopy region.*

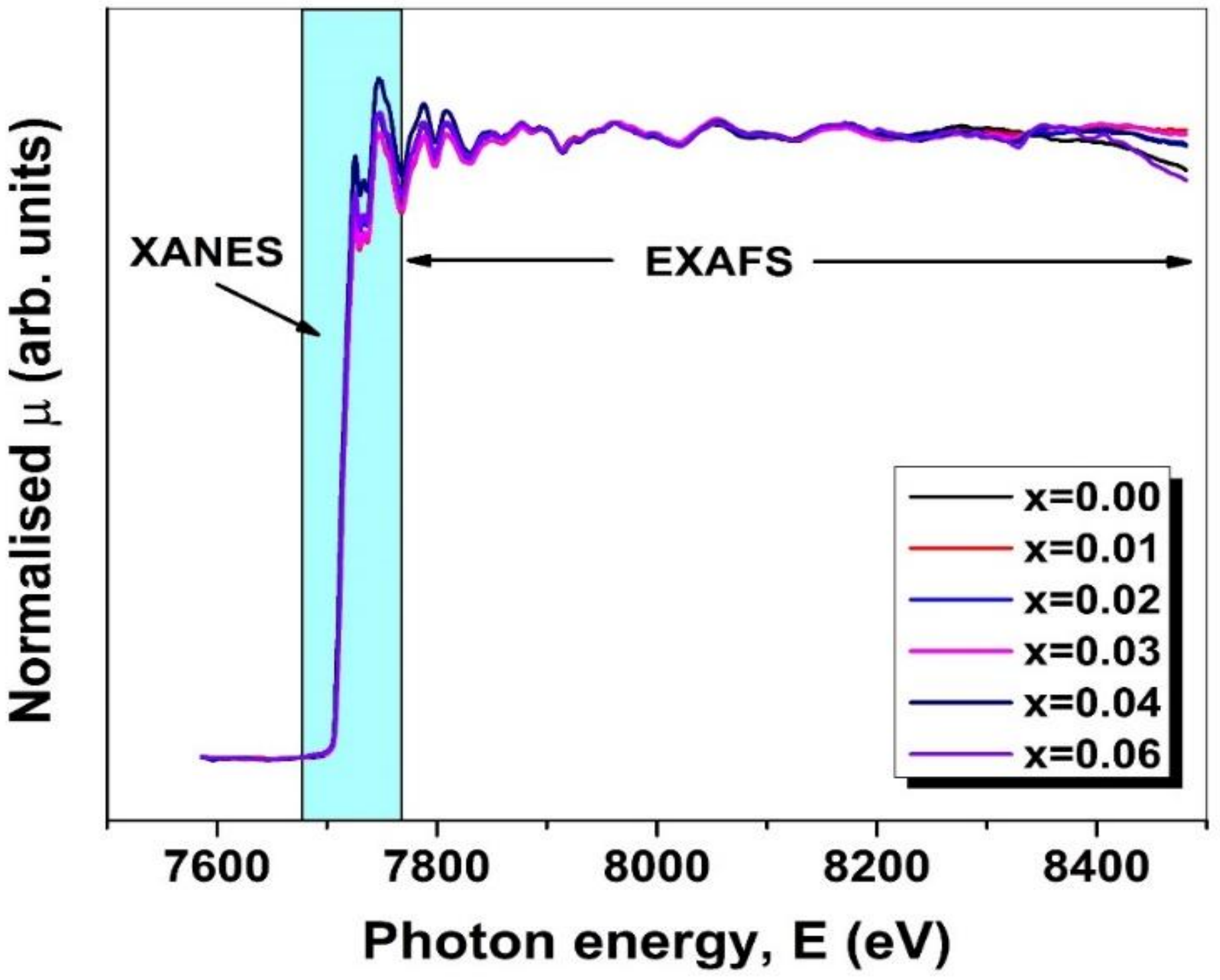

***Figure S6.*** *X-ray absorption fine structure (XAFS) spectroscopy, showing the absorption coefficient (μ) as a function of photon energy (E) of the* $TiCo_{1-x}Ni_xSb$ *(x=0, 0.01, 0.02, 0.03, 0.04, and 0.06) polycrystalline HH samples at Ti K-edge, indicating the X-ray absorption near edge structure (XANES) spectroscopy and extended X-ray absorption fine structure (EXAFS) spectroscopy region.*

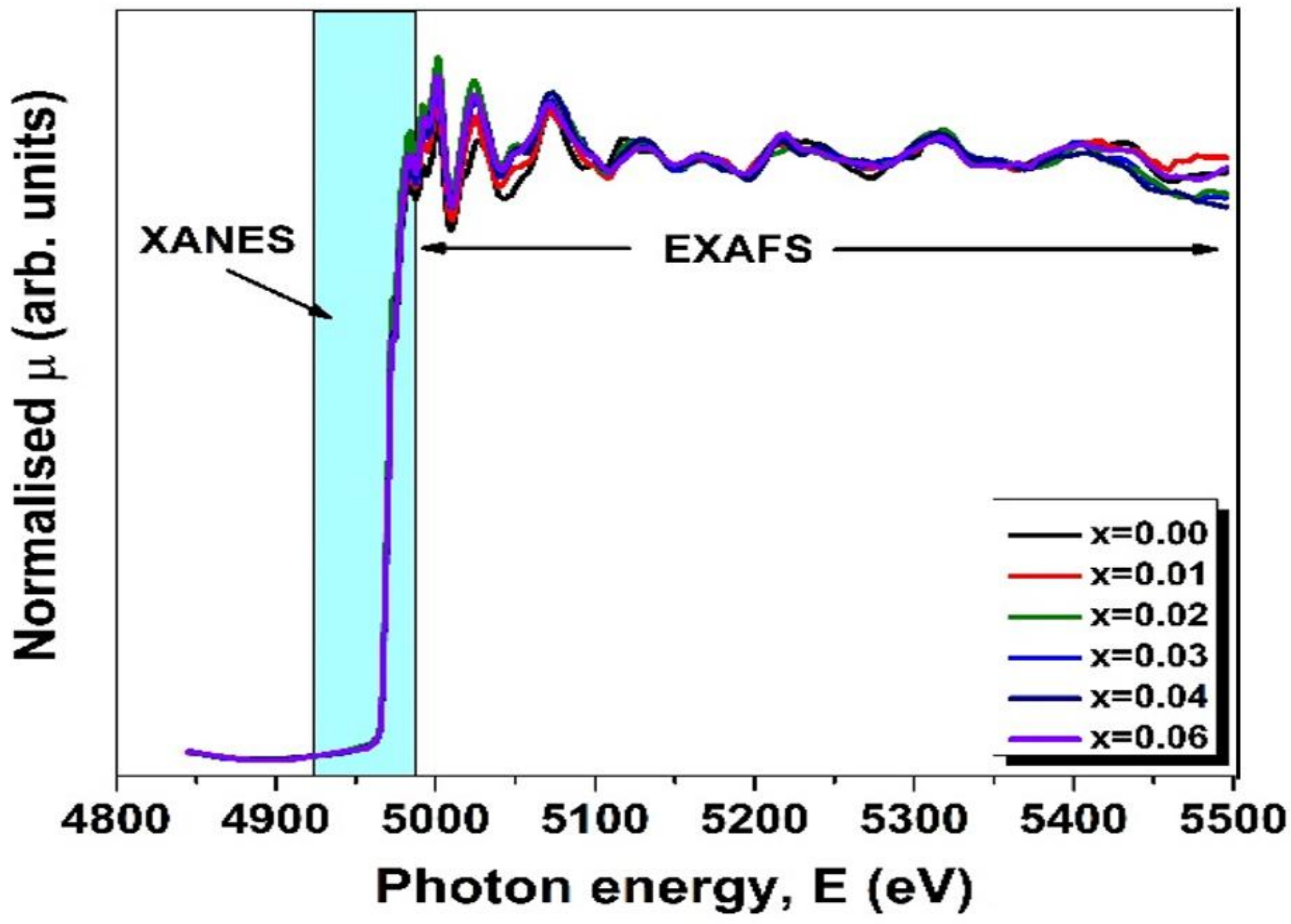

*EXAFS analysis*

The absorption coefficient, μ(E) from XAS measurements was normalized and converted to the fine structure function χ(E) [4] using the following equation,

$$\chi(E) = \frac{\mu(E)-\mu(E_0)}{\Delta\mu(E_0)} \quad \text{(S3)}$$

where $E_0$ is the threshold energy the allowed transition, $\Delta\mu(E_0)$ is the edge step, and $\mu_0(E)$ is the smooth background absorption.

Subsequently, χ(E) was transformed to k-space, χ(k), using the equation [4],

$$k = \sqrt{2m_e(E - E_0)/\hbar^2} \quad \text{(S4)}$$

where $m_e$ is the electronic mass. The $k^2$-weighted χ(k) was Fourier transformed to real space (R-space) with the ARTEMIS software package [5]. Structural fittings were performed in R-space using crystallographic information of the TiCoSb HH phase. The spectral fits employed all single scattering paths and one significant multiple scattering path with higher rank, generated from IFEFFIT integration within ARTEMIS [5]. Fitting parameters included the amplitude reduction factor ($S_0^2$), energy shift ($\Delta E_0$), change in half path length (Δr), and mean square displacement ($\sigma^2$). The number of free variables was consistently kept below the upper limit defined by the Nyquist theorem [5]. The quality of the fitting is asseced by R-factor ($R_f$) [4, 7].

***Figure S7.*** *$k^2$-weighted EXAFS spectra, χ(k), at the Co K-edge for $TiCo_{1-x}Ni_xSb$ (x = 0, 0.01, 0.02, 0.03, 0.04, and 0.06), highlighting oscillatory fine structure related to local atomic environment around Co atom.*

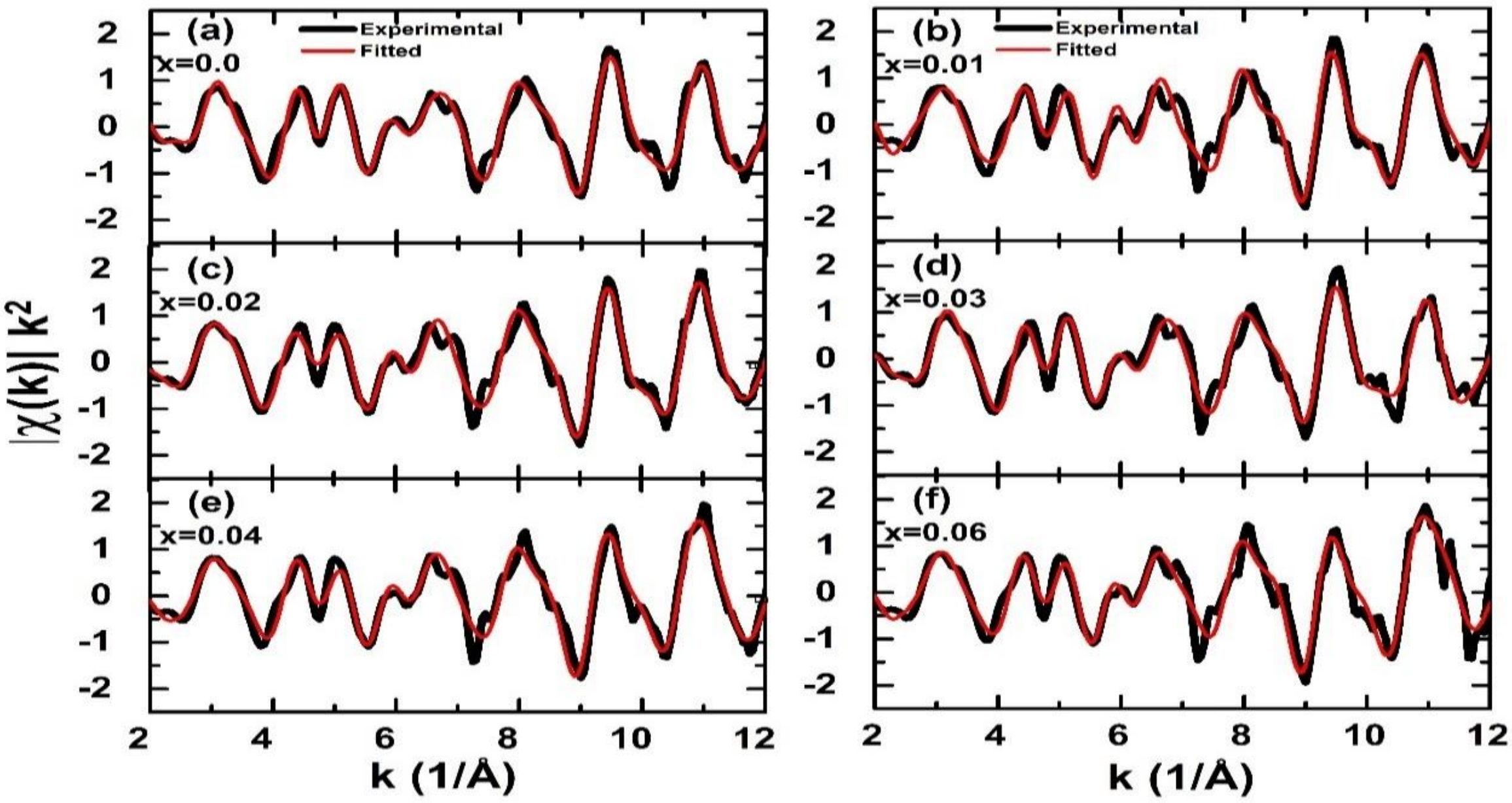

***Figure S8.*** *$k^2$-weighted EXAFS spectra, χ(k), at the Ti K-edge for $TiCo_{1-x}Ni_xSb$ (x = 0, 0.01, 0.02, 0.03, 0.04, and 0.06), highlighting oscillatory fine structure related to local atomic environment around Ti atom.*

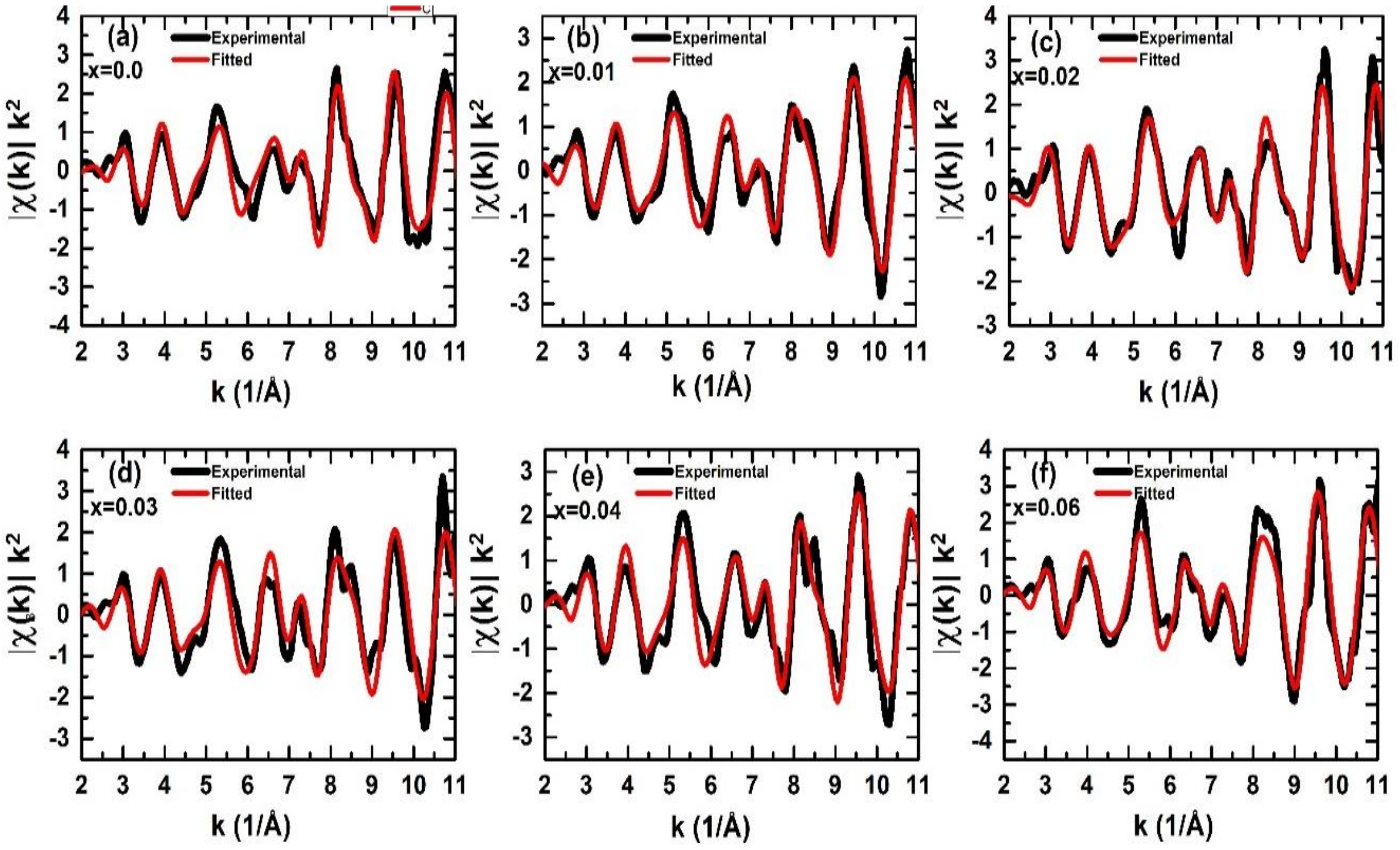

## 5. Analysis of Non-Monotonic Temperature Dependence of Electrical Resistivity (ρ(T))

Temperature dependent resistivity (ρ(T)) data of $TiCo_{1-x}Ni_xSb$ (x=0, 0.01, 0.02, 0.03, 0.04 and 0.06) synthesized samples show nonmonotonic behavior and transition from metallic to semiconducting nature for the range $0.0 \leq x \leq 0.3$. However, synthesized samples become metallic for the range $0.03 \leq x \leq 0.06$. In order to reveal the effect of scattering parameters and change in the DOS, ρ(T) data of the synthesized samples are fitted using two distinct theoretical models [8].

The high-temperature semiconducting regime of the ρ(T) data, where dρ/dT < 0, i.e., for T > 150 K, are fitted using the following equation [8],

$$\rho(T) = \rho_0^s \exp\left(\frac{-E_{act}}{2K_B T}\right). \tag{S5}$$

Where,

- $\rho_0^s$ is a pre-exponential factor, determined by the intrinsic properties of material such as carrier mobility and density of states.
- $E_{act}$ is the activation energy associated with thermally activated carrier transport in the semiconducting region.
- $K_B$ is the Boltzmann constant.

The low-temperature (T < 150 K) region of ρ(T), exhibits metallic-like behaviour (dρ/dT > 0), are fitted using the following model [8],

$$\rho(T) = \rho_0^m + \mathrm{a}T + \mathrm{b}T^2\,. \tag{S6}$$

Here:

- $\rho_0^m$ represents the residual resistivity
- The term aT accounts for electron-phonon (e-ph) scattering contributions, dominant at low to moderate temperatures.
- The quadratic term $bT^2$ reflects electron-electron (e-e) interaction effects, significant in metallic systems at low temperatures.

***Figure S9.*** *Temperature-dependent electrical resistivity [ρ(T)] of* $TiCo_{1-x}Ni_xSb$ *(x=0, 0.01, 0.02, 0.03, 0.04, and 0.06) with the fitted curve indicated by solid red and blue lines. The low-temperature metallic parts (dρ/dT > 0) are fitted using* $\rho(T) = \rho_0^m + aT + bT^2$ *(red colour line), and the high-temperature semiconducting regime (dρ/dT < 0) is fitted employing the relation* $\rho(T) = \rho_0^s \exp\left(\frac{-E_{act}}{2K_BT}\right)$ *(blue colour line).*

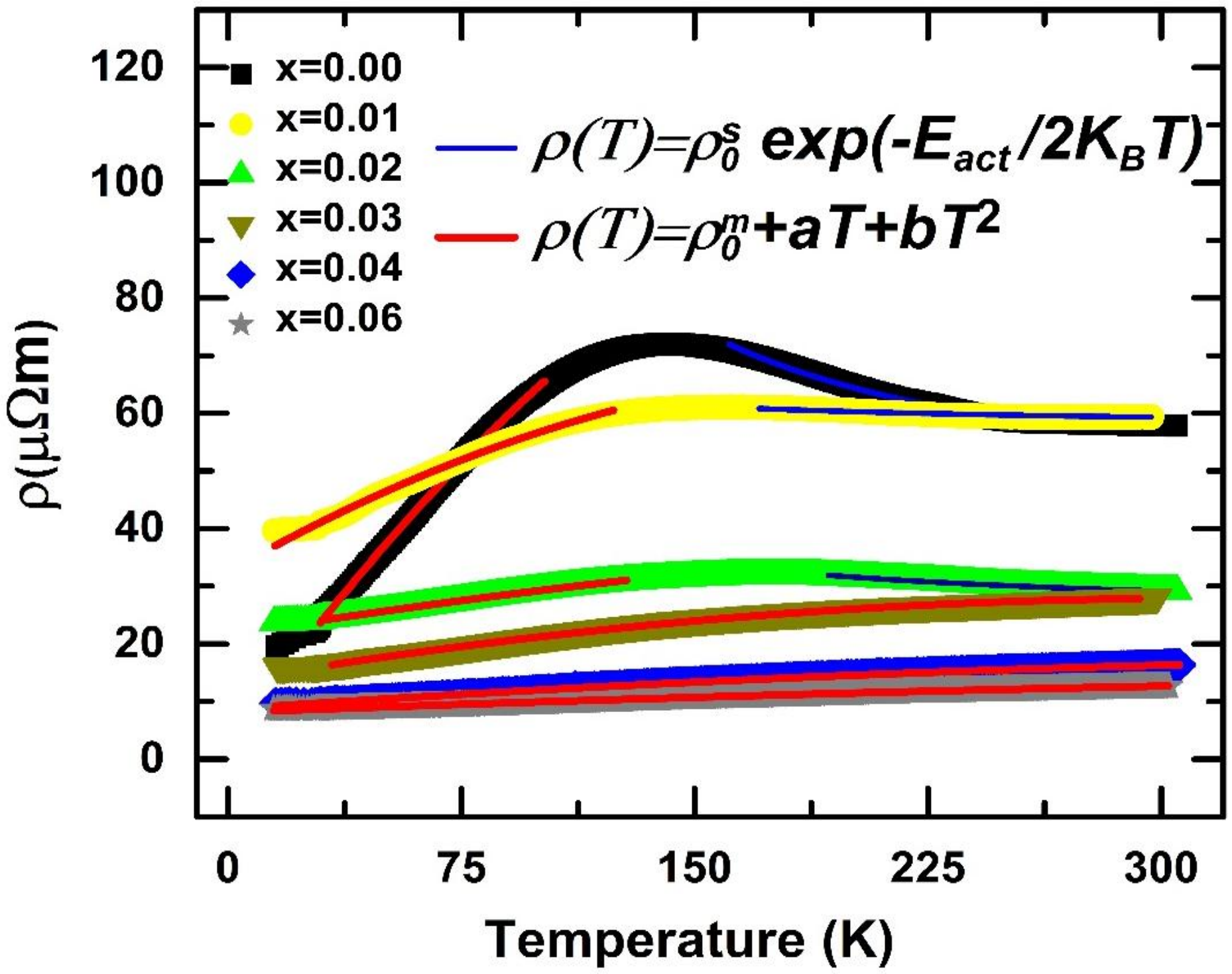